\documentclass{nature}
\usepackage{graphicx}
\usepackage{caption}
\usepackage{amsmath}
\usepackage{textcomp}
\usepackage{url}
\usepackage{lineno}
\usepackage{gensymb}
\usepackage[font=singlespacing]{caption}

\makeatletter
\let\saved@includegraphics\includegraphics
\AtBeginDocument{\let\includegraphics\saved@includegraphics}
\renewenvironment*{figure}{\@float{figure}}{\end@float}
\makeatother

\makeatletter
\long\def\@makecaption#1#2{%
  \vskip\abovecaptionskip
  \sbox\@tempboxa{#1 #2}%
  \renewcommand{\baselinestretch}{0.9}\reset@font
  \ifdim \wd\@tempboxa >\hsize
    #1 #2\par
  \else
    \global \@minipagefalse
    \hb@xt@\hsize{\hfil\box\@tempboxa\hfil}%
  \fi
  \vskip\belowcaptionskip}
\makeatother

\title{Real-space imaging of confined magnetic skyrmion tubes}

\author{M. T. Birch$^{1,2}$, D. Cort\'es-Ortu\~no$^{3}$,  L. A. Turnbull$^{1}$, M. N. Wilson$^{1}$, F. Gro\ss$^{4}$, N. Tr\"ager$^{4}$, A.~Laurenson$^{5}$, N. Bukin$^{5}$, S. H. Moody$^{1}$, M. Weigand$^{4,6}$, G. Sch\"utz$^{4}$,  H. Popescu$^{7}$, R. Fan$^{2}$, P.~Steadman$^{2}$, J.~A.~T.~Verezhak$^{8}$, G. Balakrishnan$^{8}$, J. C. Loudon$^{9}$, A.~C.~Twitchett-Harrison$^{9}$, O. Hovorka$^{3}$, H. Fangohr$^{3,10}$, F. Y. Ogrin$^{5}$ , J. Gr\"afe$^{4}$, P. D. Hatton$^{1}$}

\begin{document}

\maketitle
\begin{affiliations}
 \item Centre for Materials Physics, Durham University, Durham, DH1 3LE, UK
 \item Diamond Light Source, Didcot, OX11 0DE, UK
 \item Faculty of Engineering and Physical Sciences, University of Southampton, Southampton SO17 1BJ, UK
 \item Max Planck Institute for Intelligent Systems, 70569 Stuttgart, Germany
 \item School of Physics and Astronomy, University of Exeter, Exeter, EX4 4QL, UK
 \item Helmholtz-Zentrum Berlin f\"ur Materialien und Energie GmbH, Institut Nanospektroskopie, Kekuléstrasse 5, 12489 Berlin, Germany
 \item Synchrotron SOLEIL, Saint Aubin, BP 48, 91192 Gif-sur-Yvette, France
 \item Department of Physics, University of Warwick, Coventry, CV4 7AL, UK
 \item Department of Materials Science and Metallurgy, University of Cambridge, Cambridge, CB3 0FS, UK
 \item European XFEL GmbH, Holzkoppel 4, 22869 Schenefeld, Germany
\end{affiliations}

\pagebreak
\begin{abstract}
Magnetic skyrmions are topologically nontrivial particles with a potential application as information elements in future spintronic device architectures\cite{rosler_spontaneous_2006,iwasaki_current-induced_2013}. While they are commonly portrayed as two dimensional objects, in reality magnetic skyrmions are thought to exist as elongated, tube-like objects extending through the thickness of the sample\cite{seki_propagating_2019,milde_unwinding_2013}. The study of this skyrmion tube (SkT) state is highly relevant for investigating skyrmion metastability\cite{kagawa_current-induced_2017} and for implementation in recently proposed magnonic computing\cite{xing_skyrmion_2019}. However, direct experimental imaging of skyrmion tubes has yet to be reported. Here, we demonstrate the first real-space observation of skyrmion tubes in a lamella of FeGe using resonant magnetic x-ray imaging and comparative micromagnetic simulations, confirming their extended structure. The formation of these structures at the edge of the sample highlights the importance of confinement and edge effects in the stabilisation of the SkT state, opening the door to further investigations into this unexplored dimension of the skyrmion spin texture. 
\end{abstract}

Skyrmion states are typically stabilised by the interplay of the ferromagnetic exchange and Zeeman energies with the Dzyalohsinskii-Moriya Interaction (DMI)\cite{nagaosa_topological_2013}. In ferromagnet/heavy metal multilayer thin films, interfacial DMI is induced by symmetry-breaking spin-orbit coupling at the interface between the layers, leading to the formation of N\'eel-type skyrmions\cite{fert_skyrmions_2013,moreau-luchaire_additive_2016,woo_observation_2016}. Bulk DMI, arising due to the lack of centrosymmetry in the underlying crystal lattice, is responsible for the formation of Bloch-type skyrmions in a range of chiral ferromagnets\cite{muhlbauer_skyrmion_2009,yu_real-space_2010,yu_near_2011,seki_observation_2012,tokunaga_new_2015}. In crystals of these bulk materials the skyrmion state is typically only at equilibrium in a limited range of applied magnetic field and temperature just below the Curie temperature, $T_c$, forming a hexagonal skyrmion lattice (SkL) in a plane perpendicular to the applied magnetic field. 

\begin{figure}
\centering
\includegraphics[width=0.6\textwidth]{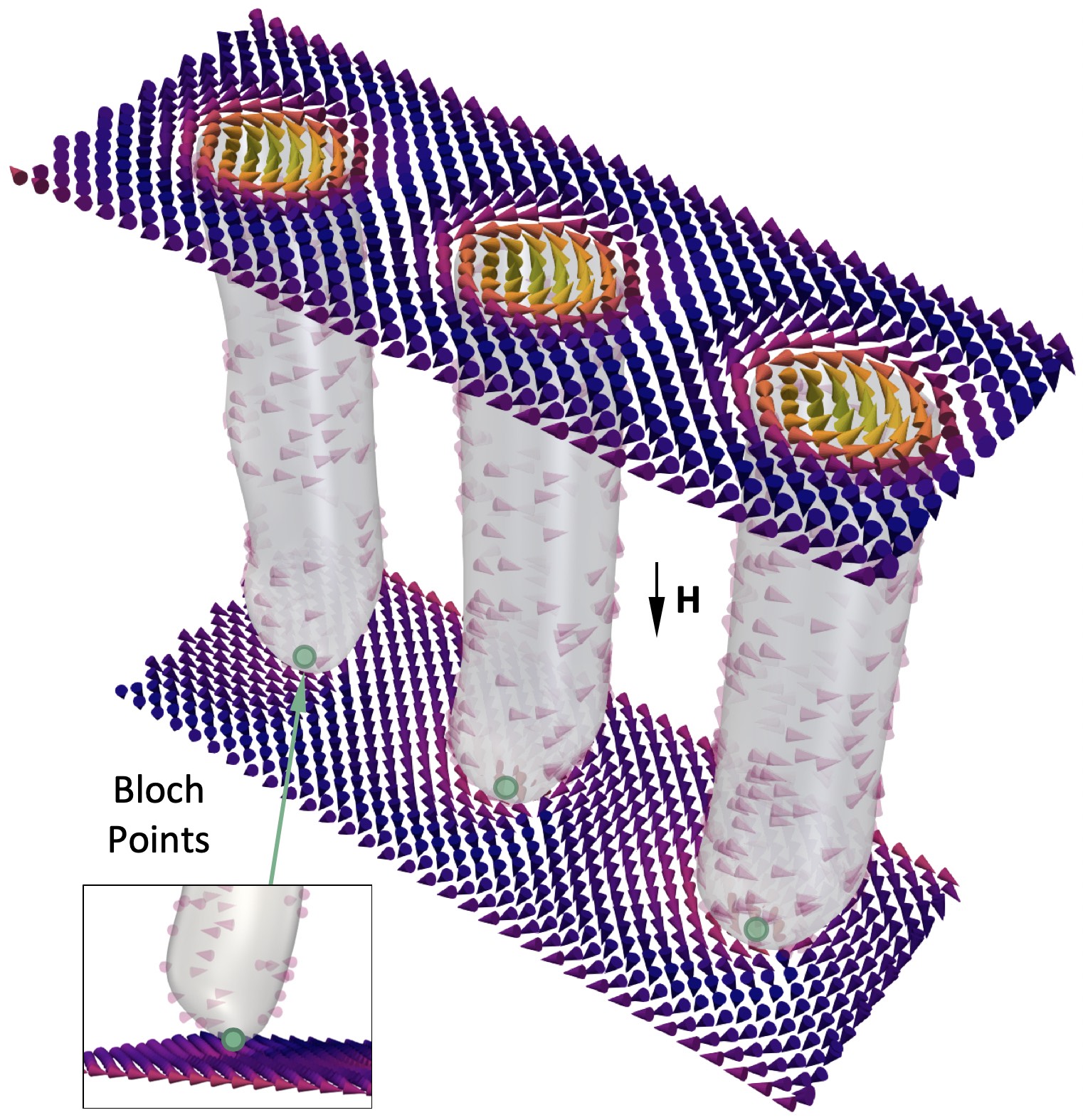}
\caption{$\vert$ \textbf{Visualisation of the skyrmion tube spin texture.} Three dimensional visualisation of three magnetic skyrmion tubes from the micromagnetic simulations presented in this paper, illustrating their extended spin structure. The inset highlights the location of the magnetic Bloch point at the end of each skyrmion tube.}
\label{fig1}
\end{figure}

The three dimensional visualisation in Fig. \ref{fig1} depicts the extended spin structure of three magnetic skyrmion tubes. The dynamics of this skyrmion tube (SkT) state play an important role in the creation and annihilation of skyrmions. For example, metastable skyrmions, which are created beyond the equilibrium thermal range by rapid field cooling\cite{karube_robust_2016}, are thought to unwind into topologically trivial magnetic states through the motion of a magnetic Bloch point located at the end of each individual skyrmion tube\cite{milde_unwinding_2013,kagawa_current-induced_2017}. Real-space observation of this dimension of the SkT state and its associated dynamics requires an in-plane magnetic field applied perpendicular to the imaging axis. Electron imaging techniques such as Fresnel Lorentz Transmission Electron Microscopy (LTEM)\cite{yu_real-space_2010,yu_near_2011}, and electron holography\cite{park_observation_2014} have been widely utilised to image magnetic skyrmions. However, due to the deflection of electron trajectories by magnetic fields, these techniques do not easily allow for the application of an in-plane magnetic field\cite{arita_development_2014}.

In magnetically sensitive x-ray based techniques, such as x-ray holography and Scanning Transmission X-ray Microscopy (STXM), the probe particles are not deflected by magnetic fields, and therefore imaging with an in-plane applied magnetic field is feasible. However, x-ray holography has seen only limited use for imaging bulk DMI skyrmions\cite{ukleev_element_2019}, while STXM instruments have previously lacked cryogenic temperature capabilities, limiting their application to observing interfacial DMI skyrmions in multilayer thin films\cite{moreau-luchaire_additive_2016,woo_observation_2016}. In this work, we utilise both x-ray holography and cryogenic STXM to image chiral spin textures in FeGe lamellae, and, with comparative micromagnetic simulations, demonstrate the first real-space observation of magnetic skyrmion tubes.

\begin{figure}
\centering
\includegraphics[width=0.7\textwidth]{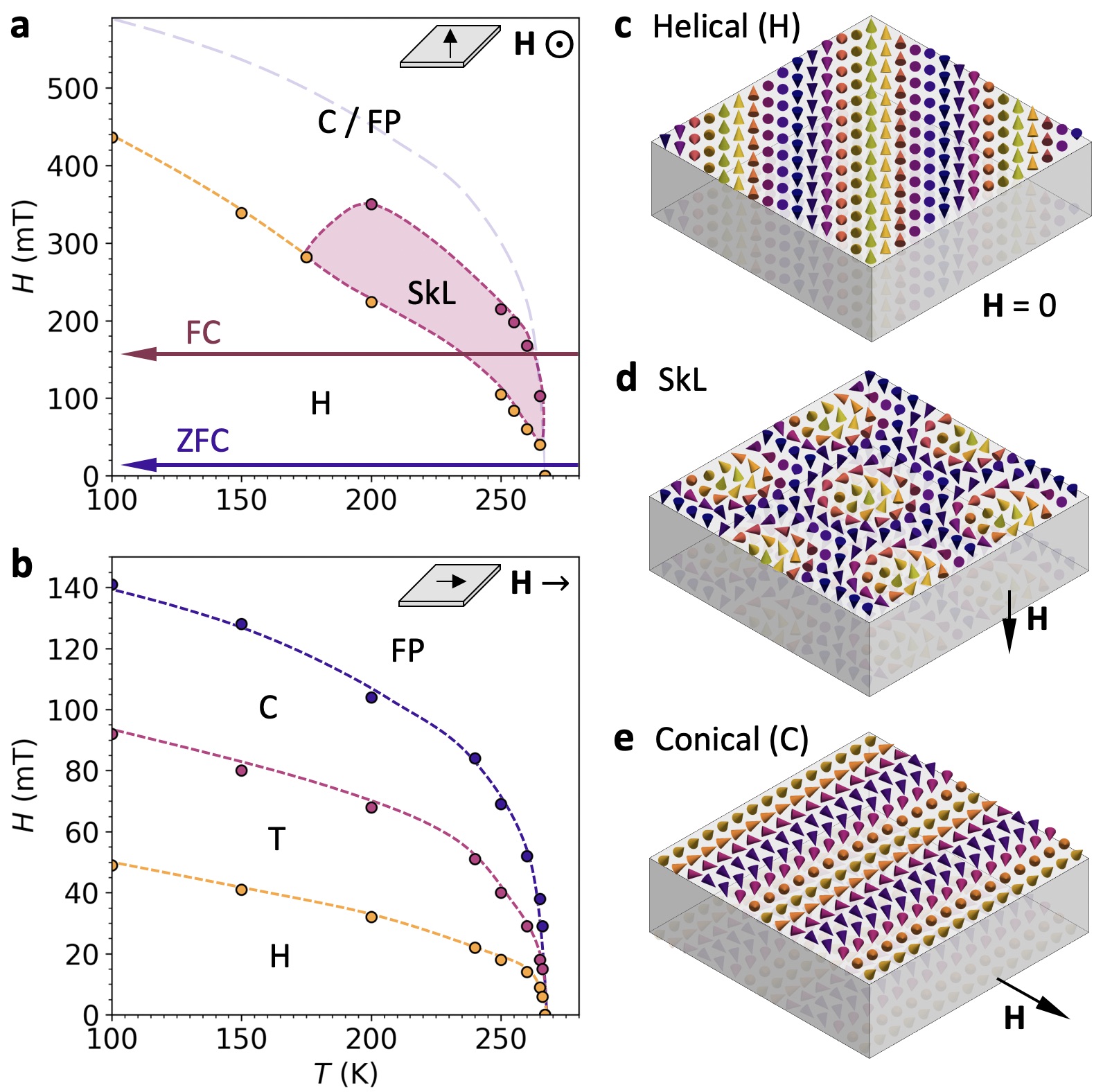}
\caption{$\vert$ \textbf{Spin textures and magnetic phase diagrams of an FeGe lamella.} \textbf{a},\textbf{b}, Phase diagrams of the $\sim$120 nm FeGe lamella for out-of-plane and in-plane applied magnetic field respectively, as determined magnetic x-ray diffraction. Schematics of each field configuration are shown as insets. In \textbf{b}, boundaries between the helical (H) and skyrmion lattice (SkL) states are displayed by yellow and magenta dots. The expected boundary between the indistinguishable conical (C) and field polarised (FP) states is estimated by the purple dashed line. In \textbf{c}, yellow, magenta and purple dots boundaries indicate the boundaries between the helical (H), helical rotation transition (T), conical (C) and field polarised (FP) states. \textbf{c}-\textbf{e}, Schematic illustrations of the spin textures as they are expected to appear in a thin lamella.}
\label{fig2}
\end{figure}

Magnetic phase diagrams of a $\sim$120 nm thick FeGe lamella (see Methods) are displayed in Fig. \ref{fig2}\textbf{a} and \textbf{b} for magnetic fields applied out-of-plane and in-plane respectively, as determined by magnetic diffraction measurements (see Methods and Supplementary Fig. S2 and S3). At low applied magnetic fields the helical state is at equilibrium, consisting of a continuous rotation of spins orthogonal to a propagation vector, as depicted in Fig. \ref{fig2}\textbf{c}. This vector lies in the plane of the lamella along a preferred axis determined by the present cubic anisotropy\cite{siegfried_spin-wave_2017}. Upon application of an out-of-plane magnetic field, the SkL state is formed, illustrated in Fig. \ref{fig2}\textbf{d}. In this field configuration, the extent of the equilibrium SkL state is greatly expanded in comparison to bulk (see Supplementary Fig. S1). This phenomenon has previously been attributed to shape anisotropy and confinement effects due to the reduced dimensionality of the sample\cite{du_highly_2014, leonov_chiral_2016,yu_observation_2013,zhao_direct_2016}. At higher out-of-plane magnetic fields, the magnetisation is expected to form the out-of-plane conical and field polarised states. However, these are indistinguishable for diffraction measurements in this field configuration.

When an in-plane magnetic field is applied, the helical state rotates as it transitions to the conical structure, which is comprised of a continuous rotation of spins at an acute angle to a propagation vector aligned parallel to the applied magnetic field. The application of an in-plane magnetic field is also expected to stabilise the in-plane SkT state. However, we found that in this field configuration the extent of the equilibrium skyrmion region was greatly suppressed, possibly entirely, as evidenced by the lack of a SkT state in Fig. \ref{fig2}\textbf{b}. This behaviour can be expected when considering that the effects of shape anisotropy and confinement, which enhance the stability of the SkL in the out-of-plane field configuration, may work to reduce the stability of the SkT state for the in-plane field arrangement. We note that due to the sample construction required for these diffraction measurements, the field of view was limited to the centre of the lamella, and therefore it was not possible to detect potential formation of a SkT state at the edges of the sample (see supplementary Fig. S1).

\begin{figure}
\centering
\includegraphics[width=\textwidth]{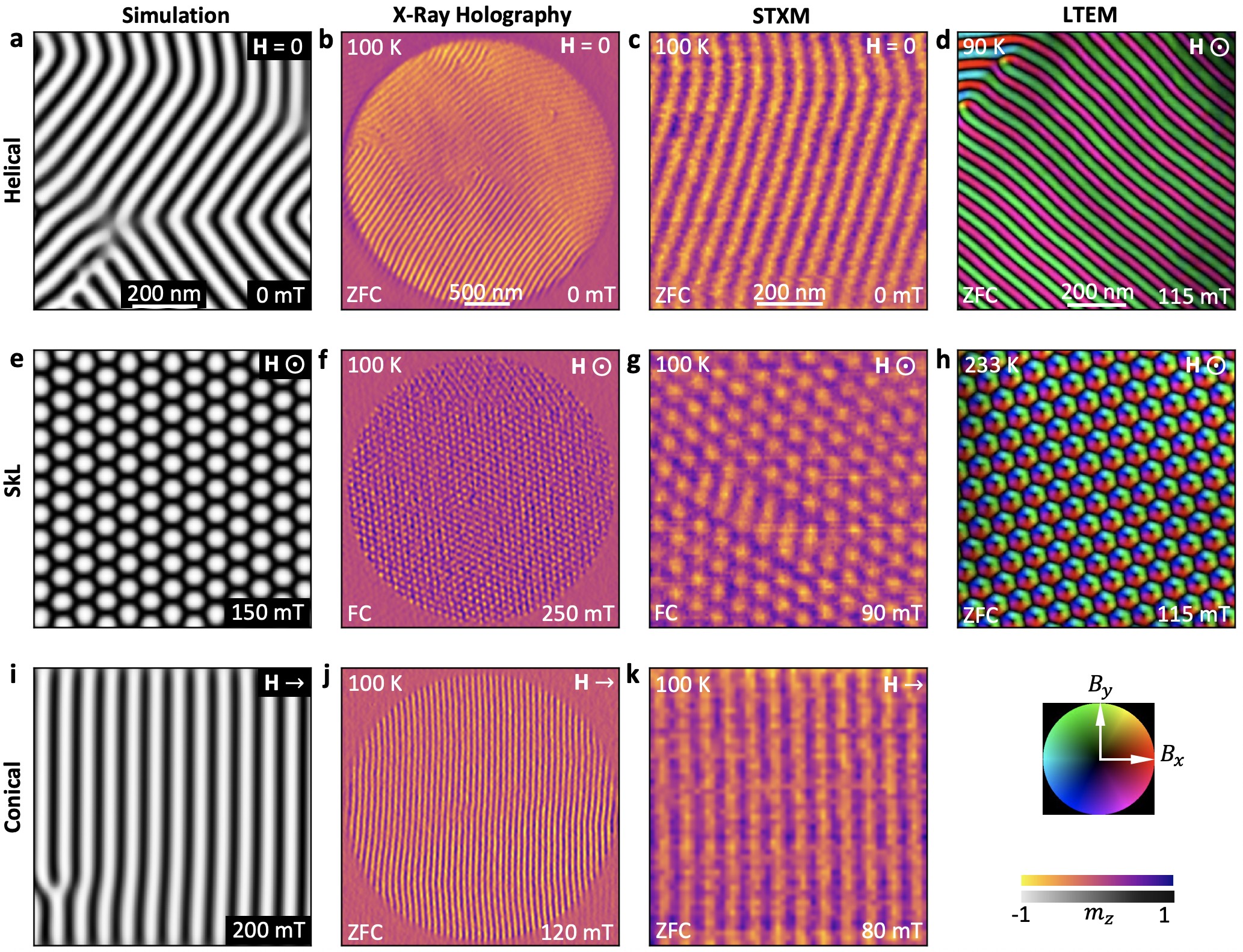}
\caption{$\vert$ \textbf{Real-space imaging of chiral spin textures.} Micromagnetic simulations, experimental x-ray holography images, STXM images and LTEM images of the (\textbf{a}-\textbf{d}) helical, (\textbf{e}-\textbf{h}) SkL and (\textbf{i}-\textbf{k}) conical magnetic spin textures. The simulation, x-ray holography and STXM images plot the normalised out-of-plane magnetisation, $m_z$, as the colour map, while the LTEM images plot the in-plane magnetic flux density as the colour map, with the direction indicated by the colour wheel at the bottom right.}
\label{fig3}
\end{figure}

Magnetic images acquired by the x-ray holography, STXM and LTEM techniques are presented in Fig. \ref{fig3}. Both x-ray imaging techniques achieve magnetic contrast by exploiting the resonant enhancement of the scattering and absorption of x-rays close to the  L$_3$ absorption edge of the magnetic Fe atoms, exhibiting a signal proportional to $m_z$, the out-of-plane component of the sample magnetisation (see Methods). In contrast, LTEM provides the in-plane components of the magnetic flux generated by the underlying magnetisation, highlighting the complimentary nature of these two techniques (see Methods). Simulated x-ray images were created from comparative micromagnetic simulations by averaging the simulated out-of-plane magnetisation $m_z$ through the thickness of the simulated spin texture, and are in excellent agreement with the experimental images (See Methods and Supplementary Fig. S5 and S6 for details). 

To achieve sufficient magnetic contrast in both the x-ray holography and STXM measurements, we found that it was necessary to maximise the ordered magnetic moment by acquiring images below 150 K. The helical state is featured in Fig. \ref{fig3}\textbf{a}-\textbf{d}, demonstrating the formation of stripe-like structures in diagonal orientations with a measured period of $\sim$ 68 nm. Figures \ref{fig3}\textbf{e}-\textbf{h} display images of the SkL state for an out-of-plane applied magnetic field, with a measured period of $\sim$ 76 nm. As no equilibrium SkL state is present at 150 K and below, we utilised field cooling to generate a metastable SkL state for the x-ray images presented in Fig. \ref{fig3}\textbf{f} and \textbf{g}. Images of the conical state under an in-plane magnetic field are displayed Figs. \ref{fig3}\textbf{i}-\textbf{k}, with a period of $\sim$ 68 nm. 

\begin{figure*}
\centering
\includegraphics[width=0.95\textwidth]{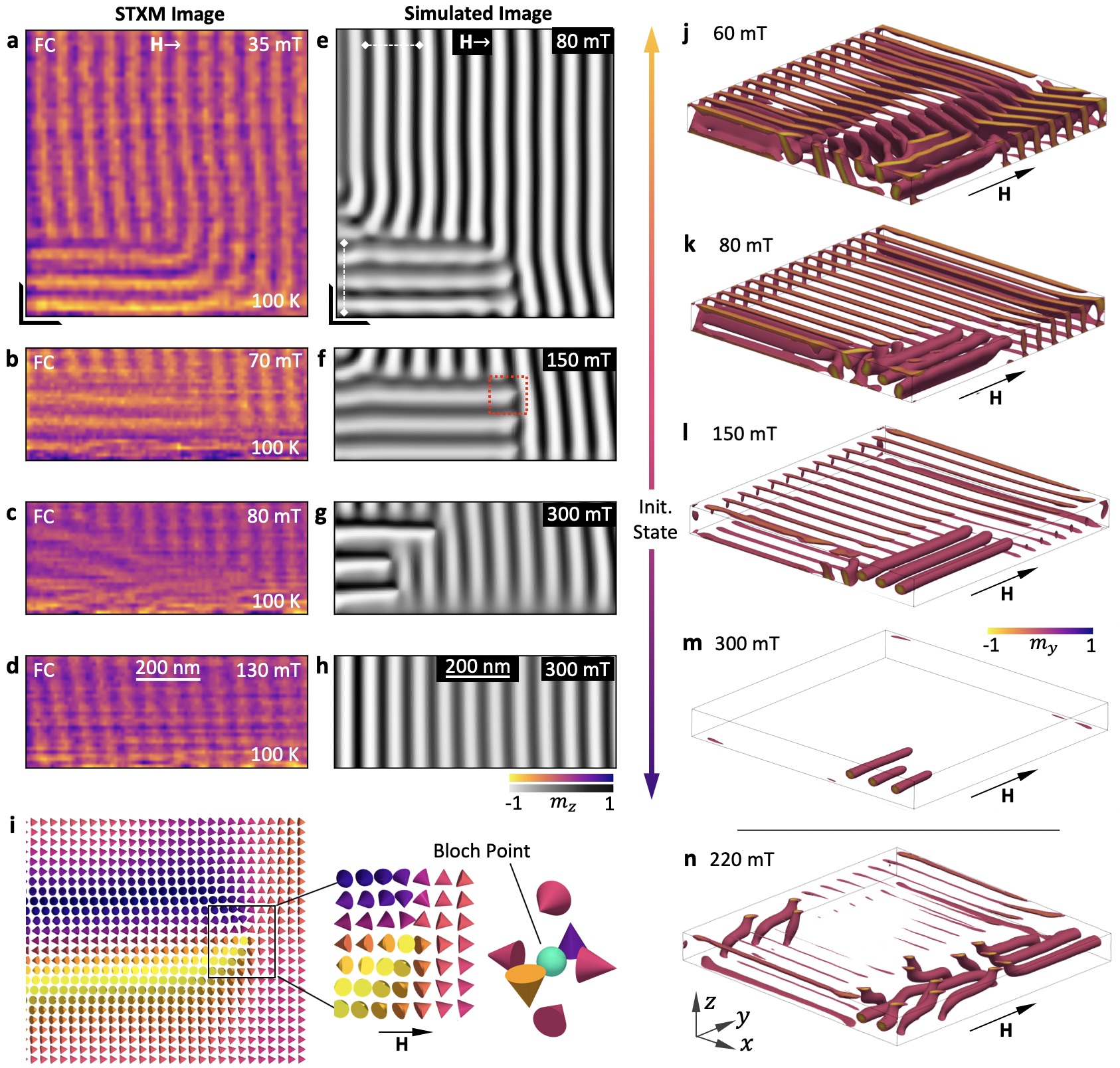}
\caption{$\vert$ \textbf{Experimental observation and micromagnetic simulations of skyrmion tubes.} \textbf{a}-\textbf{d}, Scanning transmission x-ray micrographs of the skyrmion tube spin texture embedded in the conical state observed as a function of applied in-plane magnetic field. The colourmap plots the normalised out-of-plane magnetisation averaged through the thickness of the sample, $m_z$. The black L-shape in \textbf{a} indicates the location of the corner of the sample. \textbf{e}-\textbf{h}, Simulated images of the skyrmion tube spin texture embedded in the conical state, as determined from micromagnetic simulations calculated as a function of applied in-plane magnetic field, where the colourmap plots the normalised out-of-plane magnetisation averaged through the thickness. \textbf{i}, A cross section of spins from the simulation in \textbf{f}, location shown by the red box, highlighting the presence of a magnetic Bloch point at the end of each skyrmion tube. \textbf{j}-\textbf{m}, Three dimensional visualisations of the micromagentic simulations for selected magnetic fields, obtained by plotting cells with normalised  $m_y$ between -1 and 0. \textbf{n}, Three dimensional visualisation of the skyrmion tube state achieved after a field sweep from an initially randomised state.}
\label{fig4}
\end{figure*}

After demonstrating successful x-ray imaging of chiral magnetic structures for both out-of-plane and in-plane applied magnetic fields, we investigated the possibility of observing the magnetic SkT state. Figure \ref{fig4}\textbf{a} displays a STXM image acquired after field cooling a second FeGe lamella under an applied in-plane magnetic field of 35 mT. The three horizontal stripes at the bottom of the image, which are situated in the corner of the sample (see supplementary Fig. S1 and S4), are aligned along the applied magnetic field direction, and thus have the expected appearance of the SkT spin texture embedded in the vertical stripes of the conical state. While the uppermost skyrmion tube bends directly into the conical stripes, the two lower tubes appear to bulge outward at their ends. Upon increasing the applied magnetic field, these skyrmion tubes decrease in length, before being annihilated by the conical state at 130 mT, as shown in Figs. \ref{fig4}\textbf{b}-\textbf{d}. 

To validate the identification of this structure as the SkT state, we performed supporting micromagnetic simulations. The simulation was initialised by relaxing a state consisting of three paraboloid-shaped skyrmion tube precusors at a range of in-plane magnetic fields, with the state at 150 mT showing the closest agreement to experiment (see Supplementary Fig. S7). The magnetic field was then varied to explore the field-dependent behaviour of the simulated SkT state. Selected simulated images are displayed in Figs. \ref{fig4}\textbf{e}-\textbf{h}, showing remarkable agreement to the presented experimental micrographs. In Fig. \ref{fig4}\textbf{e}, at 90 mT, the end of the uppermost skyrmion tube curves into the conical state, while the two lower tubes appear to bulge at the end, replicating the behaviour observed in the experimental image Fig. \ref{fig4}\textbf{a}. A cross section through the end of one of these simulated tubes is shown in Fig. \ref{fig4}\textbf{i}, highlighting the presence of a magnetic Bloch point. The image in Fig. \ref{fig4}\textbf{a} may therefore represent the experimental observation of a magnetic Bloch point structure at the end of a skyrmion tube. 

Selected three dimensional visualisations of the simulations are displayed in Figs. \ref{fig4}\textbf{j}-\textbf{m}. The additional surface structures in Fig. \ref{fig4}\textbf{j}-\textbf{l} are chiral edge twists in the conical state at the sample boundary\cite{meynell_surface_2014}. At decreasing magnetic fields, the skyrmion tubes branch into the helical state and expand to touch the surfaces of the simulated sample, establishing partial skyrmion tube edge states shown in Fig. \ref{fig4}\textbf{j}. Upon increasing the magnetic field, the skyrmion tubes decrease in length, as seen in the experimental images. Despite the excellent qualitative agreement of the experimental and simulated images exhibited in Fig. \ref{fig4}, we note that in the simulation the SkT state exists over a higher magnetic field range in comparison to the experimental observations. However, this can be attributed to two factors. Firstly, micromagnetic simulations are not able to incorporate thermal effects, and are effectively performed at 0 K, whereas the experimental images were acquired at 100 K. Secondly, the lateral extent of the simulated sample is smaller than the measured lamella, exaggerating the effects of the demagnetising field.

We performed an additional simulation commencing from a randomly initialised helical state at 0 mT, and found that the SkT state was also stabilised during an in-plane magnetic field sweep, as depicted by the visualisation in Fig. \ref{fig4}\textbf{n} (further visualisations in Supplementary Figs. S8 and S9). In contrast to the previous simulations, the ends of the tubes curve to touch the upper and lower faces of the sample. Such edge states may be energetically favourable in comparison to the formation of a magnetic Bloch point. Previous studies have demonstrated that the SkL state has improved stability at the sample boundaries for out-of-plane magnetic fields\cite{du_edge_2015}. Our results suggest that the stability of the SkT state is similarly enhanced at the sample edge for in-plane magnetic fields. This may also explain why no SkT state was observed in the in-plane magnetic phase diagram in Fig. \ref{fig2}\textbf{b}, where the field of view was restricted to the centre of the FeGe lamella.

\begin{figure*}
\centering
\includegraphics[width=0.5\textwidth]{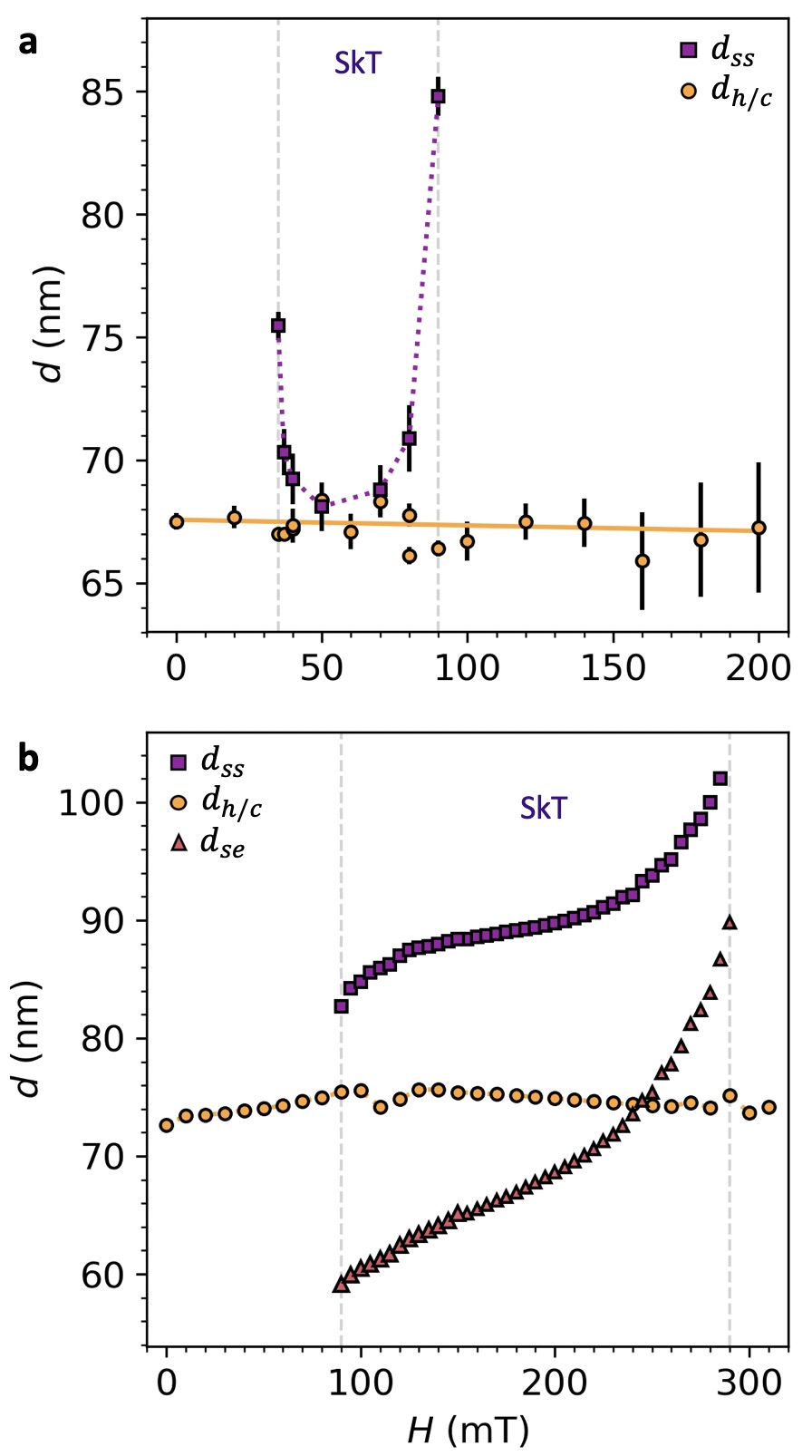}
\caption{$\vert$ \textbf{Spacing of the skyrmion tube state.} \textbf{a},\textbf{b}, The skyrmion-skyrmion tube spacing, $d_{ss}$, the skyrmion-edge distance $d_{se}$, and helical/conical state period, $d_{h/c}$, plotted as a function of applied magnetic field for the experimental and simulated images respectively.}
\label{fig5}
\end{figure*}

The helical/conical period, $d_{h,c}$, and the distance between each skyrmion tube, $d_{ss}$, were extracted from the experimental and simulated data, and are plotted as a function of applied magnetic field in Fig. \ref{fig5}\textbf{a} and \textbf{b} respectively. The skyrmion-edge distance $d_{se}$ for the simulated data is also plotted. In both the simulation and experiment, $d_{h,c}$ was found to remain approximately constant across the full range of applied magnetic field. In contrast, the simulated $d_{ss}$ and $d_{se}$ increase with applied magnetic field, exhibiting skyrmion-skyrmion and skyrmion-edge interactions which are attractive at low fields, and repulsive at higher fields (see Supplementary Fig. S11 for details), in agreement with studies of such interactions for the out-of-plane SkL state\cite{leonov_three_2016,du_interaction_2018}. At low and high fields, $d_{ss}$ becomes respectively smaller and larger than the skyrmion radius. In both instances, the magnetisation of the SkT spin texture necessarily distorts, in a manner functionally similar to the distortion of the helical state into a chiral soliton lattice\cite{okamura_emergence_2017} (see Supplementary Fig. S12). 

While the experimental behaviour of $d_{ss}$ is not exactly replicated by the simulations, its value is nevertheless highly variable, ranging from 67 nm to 86 nm, in contrast to the constant value of $d_{h,c}$. This provides strong evidence that the observed SkT spin texture is distinct from these topologically trivial magnetic states. We suggest that the discrepancy may be due to the formation of the SkT structure by field cooling, producing a pinned metastable state which first relaxes over the first few field increments, before displaying the expected increase in $d_{ss}$ at higher applied fields. Such pinning effects are not considered in the simulation model.

In conclusion, we have demonstrated direct imaging of magnetic skyrmion tubes utilising x-ray microscopy and comparative micromagnetic simulations. This observation confirms in real-space, for the first time, the extended nature of the magnetic skyrmion along the applied magnetic field direction. The field-evolution of the SkT state, and its location at the sample boundary in both the experiment and simulation, highlights the importance of confinement and boundary effects in the formation of this magnetic state and the emergent skyrmion-skyrmion and skyrmion-edge interactions. Experimental realisation of skyrmion tubes and their associated Bloch point structures opens the door to further studies of this unexplored dimension of the skyrmion spin texture and its associated dynamic phenomena.

\begin{methods}

\subsection{Sample Preparation.} Single crystals of FeGe were grown by the chemical vapour transport technique. Two grams of prepared FeGe powder along with 2 mg/cc of the iodine transporting agent was used for the growth, with the source maintained at 450 \degree C and a temperature gradient of 50 \degree C across the length of the tube, over a period of 1-2 weeks. Several single crystals with dimensions of 1.5 $\times$ 1.5 $\times$ 1.5  mm$^3$ were obtained at the colder end. From one single crystal crystal, two lamellae, of thickness $\sim$120 nm, with the [112] crystal direction as the plane-normal direction, were prepared via a lift-out method using a focused ion beam system (FEI Helios Nanolab). Using a in-situ micromanipulator, one lamella was fixed by Pt deposition over a 3 \textmu m aperture in a Si$_3$N$_4$ membrane coated with 600 nm of Au for the magnetic diffraction measurements. A reference slit of width $\sim$20 nm and length 6 \textmu{}m was cut 3.5 \textmu{}m from the centre of the sample aperture, again using the focused ion beam, for x-ray holography. The second lamella, also $\sim$120 nm thick, was attached to a standard Cu grid with Pt deposition, and ion-milled into an L-shape for STXM measurements. See Supplementary Fig. S1 for sample images.

\subsection{Magnetic X-Ray Diffraction.} Resonant magnetic x-ray diffraction measurements were performed with the RASOR diffractometer at Diamond Light Source, and the COMET instrument at Synchrotron SOLEIL. Sample cooling was achieved by a He cryostat and the applied magnetic field was controlled by varying the arrangement of four permanent magnets. The coherent x-ray beam was directed through the sample aperture and reference slit, and the resultant diffraction pattern captured by a CCD placed downstream of the sample. The magnetic signal was maximised by tuning the x-ray energy to the L$_3$ Fe absorption edge, at $\sim$708 eV, by measuring an x-ray magnetic circular dichroism (XMCD) spectrum (see Supplementary Fig. S2 for a schematic and details).

\subsection{X-Ray Holography.} Using the same experimental setup as in the magnetic x-ray diffraction measurements, magnetic x-ray holography was performed at Synchrotron SOLEIL. For each holographic image, two long exposure diffraction patterns were recorded with opposite circular polarisation of the incident x-ray beam. Holographic reconstruction of the magnetic sample image was then performed using the HERALDO technique, by subtracting the two exposures, applying a linear differential filter, and finally performing a Fourier transform (see Supplementary Fig. S2). 

\subsection{Scanning Transmission X-ray Microscopy.} Scanning transmission microscopy measurements were performed at the MAXYMUS instrument at BESSY II. With the sample mounted inside the microscope, cooling was achieved by a He cryostat and the applied magnetic field was controlled by varying the arrangement of four permanent magnets. Care was taken to reduce the vibrations from the cryostat to ensure successful low temperature imaging of the sample. The x-ray beam was focused to a 22 nm spot size using a Fresnel zone plate and order separation aperture. This focused beam, once again with an x-ray energy of 708 eV, was then rastered across the sample pixel by pixel. By exploiting the effects of XMCD at the resonant x-ray energy, the transmission of the sample at each point was measured to form an image of the magnetic contrast (see Supplementary Fig. S2).

\subsection{Lorentz Electron Transmission Microscopy.} LTEM measurements were performed on a comparable FeGe lamella using an FEI Tecnai F20 transmission electron microscope operated at an acceleration voltage of 200 kV and equipped with a field-emission electron gun. Images were acquired using a Gatan imaging filter (GIF) and recorded on a 1024 $\times$ 1024 pixel charge-coupled device (CCD). Pairs of images with equal and opposite defoci were acquired at each temperature and the projected magnetic flux density was calculated from these using the transport of intensity equation (see Supplementary).

\subsection{Micromagnetic Simulations} Simulations of the various magnetic configurations observed in the experiments were performed using the micromagnetic code OOMMF\cite{OOMMF} and the data was processed using the OOMMFPy library, available online at 10.5281/zenodo.2611194. The simulated system was specified with dimensions $1000\ \text{nm} \times1000\ \textrm{nm} \times100\ \textrm{nm}$, using finite difference cells with a volume of $4\,\textrm{nm}^{3}$, and magnetic parameters of FeGe. We describe the FeGe system using the energy functional of a chiral magnet with symmetry class $T$, which reads
\begin{equation}
%\begin{split}
    E = \int_{V} \text{d}V \bigg\{
        A \sum_{\alpha=x,y,z} \left(\nabla m_{\alpha} \right)^{2}
        + D \mathbf{m} \cdot \left( \nabla \times \mathbf{m} \right)
       -M_{\text{s}} \mathbf{m} \cdot \mathbf{B}_{\text{a}}
        - \frac{M_{\text{s}}}{2} \mathbf{m}\cdot\mathbf{B}_{\text{d}} \bigg\},
%\end{split}
\end{equation}
\noindent where $\mathbf{m}$ is the normalised magnetisation, $A=8.78\, \text{pJ m}^{-1}$ is the exchange constant, $M_{\text{s}}=384\, \text{kA m}^{-1}$ is the saturation magnetisation, $D=1.58\, \text{mJ m}^{-2}$ is the DMI constant, $\mathbf{B}_{\text{a}}$ is the applied field and $\mathbf{B}_{\text{d}}$ is the demagnetising field. The energy minimisation of a specified initial state was performed using OOMMF's conjugate gradient method. The same minimisation technique was applied to reach equilibrium states at each step of the simulated field sweeps (see Supplementary Information for details). Three dimensional visualisations for all the simulations can be viewed in Supplementary Videos 1-4. 

\subsection{Data Availability}
Simulation data utilised to produce the presented figures in this paper are available at the following doi: 10.5281/zenodo.3364609.    

\end{methods}

\begin{addendum}
 \item We give thanks for the assistance of M. Sussmuth at Diamond Light Source. We acknowledge the support of Diamond Light Source for time on beamline I10 under proposal SI20866-2, as well as experiment time at SOLEIL (proposal 20180679) and BESSY II (proposal 181-06589ST). This work was financially supported by two Engineering and Physical Sciences Research Council grants: EP/M028771/1, and the UK Skyrmion Project Grant, EP/N032128/1. M.N.W. acknowledges the support of The Natural Sciences and Engineering Research Council of Canada (NSERC). 

 \item[Author Contributions] M.T.B., L.A.T., M.N.W., and P.D.H. conceived the project and designed the experiments. J.A.T.V. and G.B. produced the FeGe single crystal. M.T.B., L.A.T., N.B. and A.C.T-H. manufactured the lamella samples. M.T.B., L.A.T., S.H.M., R.F. and P.S. performed the magnetic diffraction measurements at Diamond Light Source. M.T.B., L.A.T., A.L., H.P., and F.Y.O. performed the magnetic holography experiments at SOLEIL. M.T.B., L.A.T., M.N.W., F.G., N.T., M.W. and J.G. performed the STXM measurements at BESSY II. J.C.L. and A.C.T-H acquired the LTEM data. D.C-O., O.H. and H.F. performed the micromagnetic simulations. M.T.B and D.C-O. wrote the paper. All authors discussed the results and commented on the manuscript.
 \item[Competing Interests] The authors declare that they have no competing financial interests.
 \item[Correspondence] Correspondence and requests for materials should be addressed to P. D. Hatton \newline (email: p.d.hatton@durham.ac.uk).
\end{addendum}

\pagebreak
\section*{\Huge Supplementary Information}
\setcounter{figure}{0}
\renewcommand\figurename{Figure S}
\makeatletter
\def\fnum@figure{\figurename\thefigure}
\makeatother

\section*{Sample Preparation and Characterisation}
The FeGe single crystals were grown by the chemical vapour transport method with iodine as the transport agent, as described by Richardson\cite{richardson_partial_1967}. We performed initial magnetometry measurements to characterise the bulk single crystal FeGe sample, displayed in Fig. S1\textbf{a} and \textbf{b}. The $T_c$ of the sample, defined as the point of greatest slope in $M$, was found to be 280.5 K, as determined by the magnetisation vs temperature data in Fig. S1\textbf{a}. The thin lamella samples in the main text were observed to have a $T_c$ of 270 K. This difference is likely due to two factors: a temperature offset of the lamella sample from the temperature sensor during the x-ray measurements, or a reduction in $T_c$ due to ion implantation damage to the lamella during the fabrication process. A magnetic phase diagram of the crystal is shown in Fig. S1\textbf{b}, highlighting the limited extent of the SkL state in the bulk when compared to the out-of-plane field lamella phase diagram in Fig. 1\textbf{a} in the main text. The phase diagram colourmap plots the AC susceptibility measured during field sweeps after zero field-cooling at each temperature. A region with a characteristic dip in the AC signal is a well known indicator of the SkL state.

\begin{figure}
\centering
\includegraphics[width=0.7\textwidth]{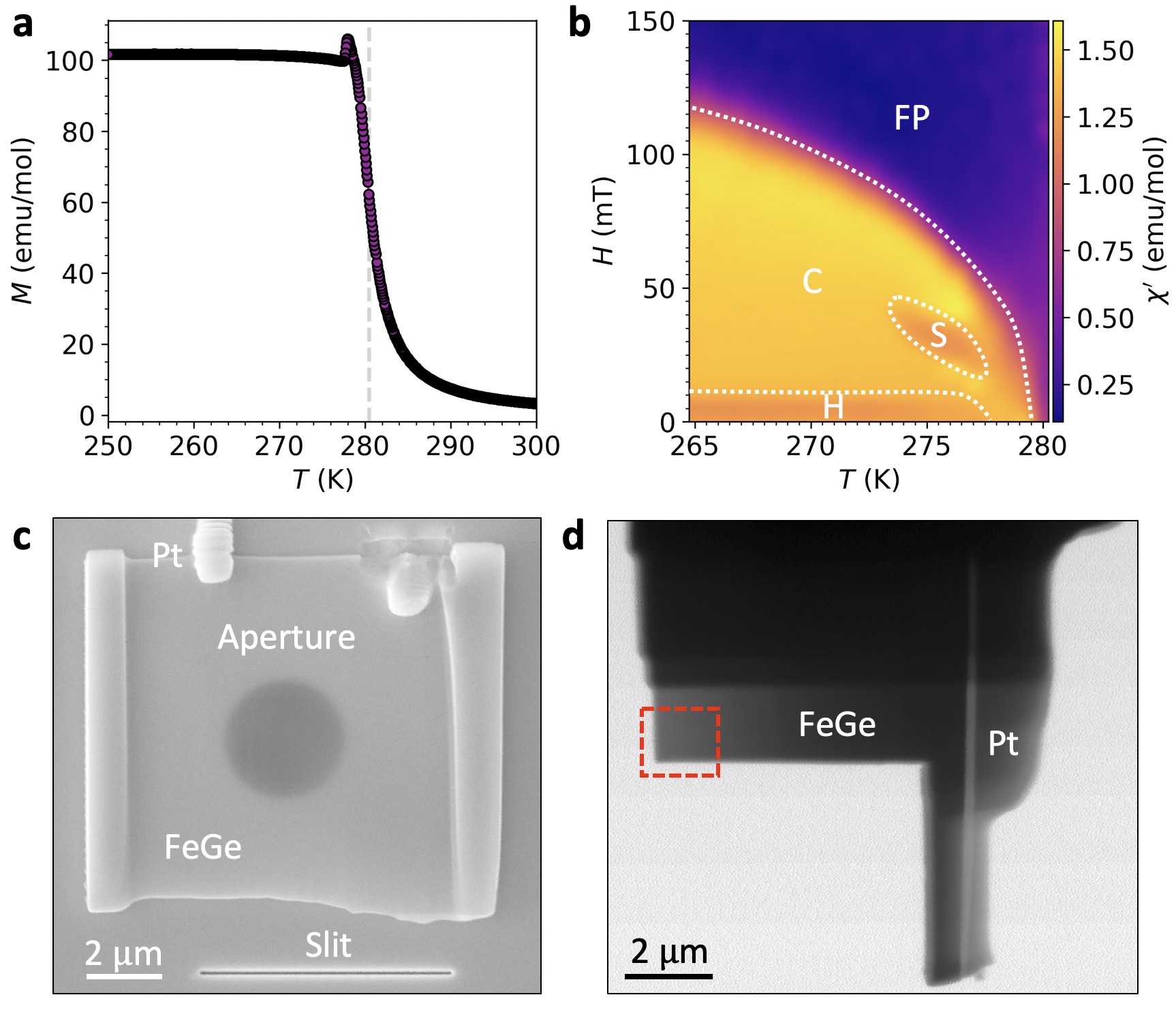}
\caption{$\vert$ \textbf{Bulk and thin lamella FeGe samples.} \textbf{a}, A plot of magnetisation against temperature as measured for the bulk FeGe single crystal. The grey dotted line marks the determined T$_c$ of $\sim$ 280.5 K. \textbf{b}, The magnetic phase diagram of the bulk FeGe sample, as determined by AC susceptibility measurements. The extent of the helical (H), conical (C), skyrmion lattice (S) and field polarised (FP) states are labelled. \textbf{c}, Scanning electron micrograph of sample 1. \textbf{d}, Scanning transmission x-ray micrograph of sample 2. The red box indicates the location of the observed magnetic skyrmion tubes.}
\label{figS1}
\end{figure}

Microscopy images of the thin lamella samples of FeGe investigated in the main text are shown in Fig. S1\textbf{a} and \textbf{b}. Sample 1, shown in Fig. S1\textbf{c}, consists of a $\sim$120 nm thick lamella fabricated by focused ion beam. This lamella was placed over a 3 \textmu m aperture in a Si$_4$N$_3$ membrane, which was sputter coated on the reverse side with $\sim$600 nm of Au, and fixed in position by a single Pt deposition weld. A 20 nm thick reference slit was milled 3.5 \textmu m from the sample aperture. The Au layer serves to block the majority of the incoming x-ray beam, leaving only scattering from the sample aperture and the reference slit incident on the CCD. Sample 2, shown in Fig. S1\textbf{d}, consists of a $\sim$120 nm thick lamella again fabricated by focused ion beam. The lamella was further ion milled into the L-shape shown in the sample image. The contrast in the image demonstrates that the bottom left corner of the sample, highlighted by the dashed red rectangle, is the thinnest part of the sample. The SkT spin texture imaged in the main text was located in this corner. 

\section*{Experimental Setups}
The experimental setups for the magnetic x-ray diffraction, x-ray holography and STXM are shown in Fig. S2\textbf{a} and \textbf{b}, and are described in the methods section of the main text. Further details of the x-ray holography\cite{popescu_comet_2019,duckworth_magnetic_2011} and STXM\cite{follath_the_2010,nolle_unique_2012} techniques can be found in previous work. Figure S2\textbf{c} displays the x-ray absorption spectrum measured on sample 2. The sample absorption is resonantly enhanced when the energy of the incident x-rays are tuned to the Fe L$_3$ and L$_2$ edges\cite{blume_polarization_1988}. A difference in absorption between left and right circularly polarised x-rays indicates x-ray magnetic circular dichroism (XMCD), and this effect is exploited to achieve contrast in magnetic STXM imaging. In order to achieve magnetic contrast at these energies without suffering from spectral compression due to the increased absorption at the resonant edge, we found that the thickness of the FeGe sample was required to be less than 150 nm. Details of the transport of intensity equation analysis used to produce the LTEM images can be found in previous work\cite{Beleggia2004}.

\begin{figure}
\centering
\includegraphics[width=0.6\textwidth]{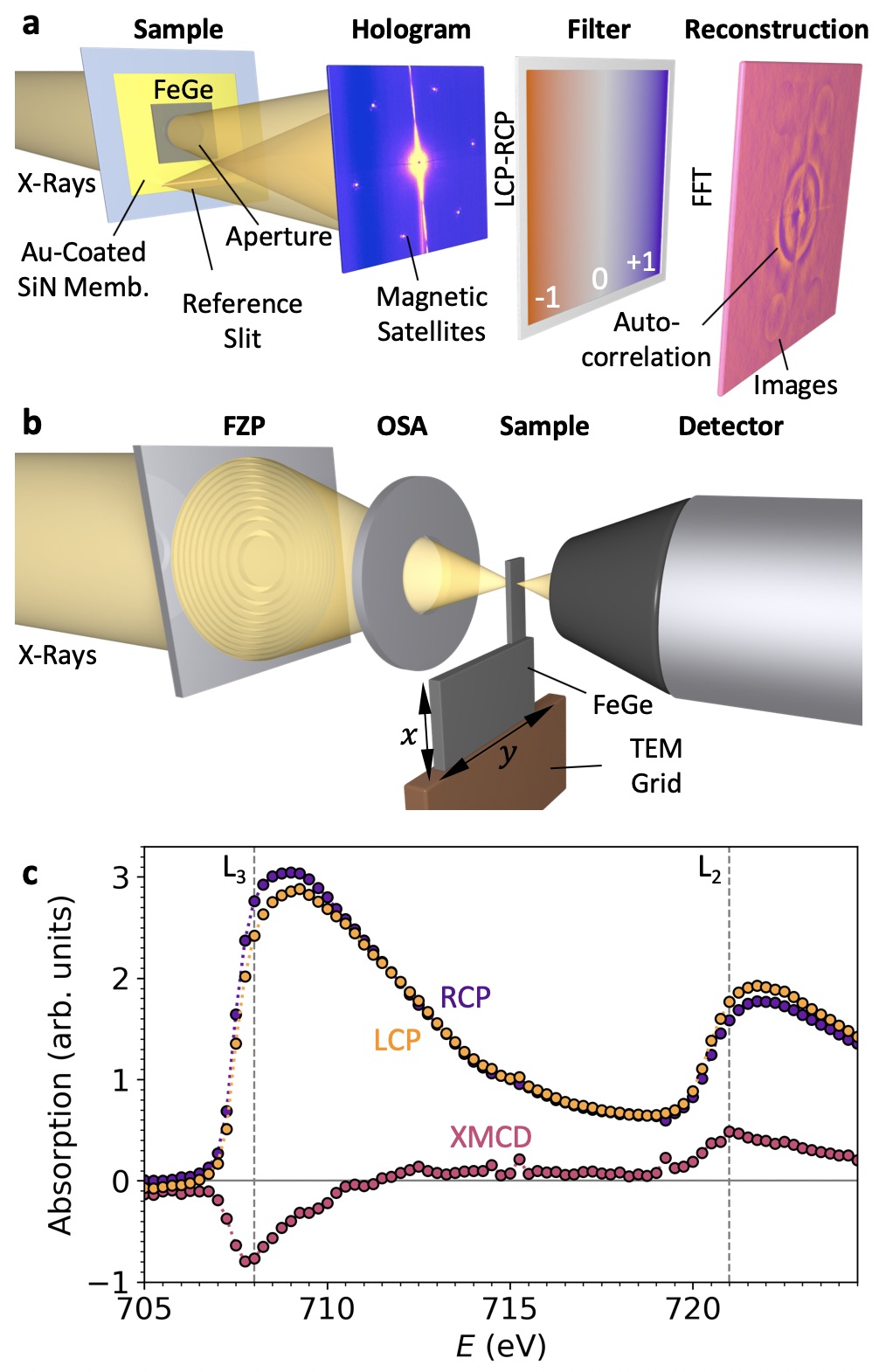}
\caption{$\vert$ \textbf{Experimental setup diagrams and x-ray spectra.} \textbf{a},\textbf{b}, Schematic illustrations of the magnetic x-ray diffraction and holographic imaging experiments (\textbf{a}), and the scanning transmission microscopy setup (\textbf{b}). \textbf{c}, Absorption spectrum of the Fe L$_3$ and L$_2$ edges, measured for both left (LCP) and right (RCP) circularly polarised x-rays. The difference is the XMCD spectrum, also displayed.}
\label{figS2}
\end{figure}

\section*{Determination of Magnetic Phase Diagrams}
\begin{figure}
\centering
\includegraphics[width=0.8\textwidth]{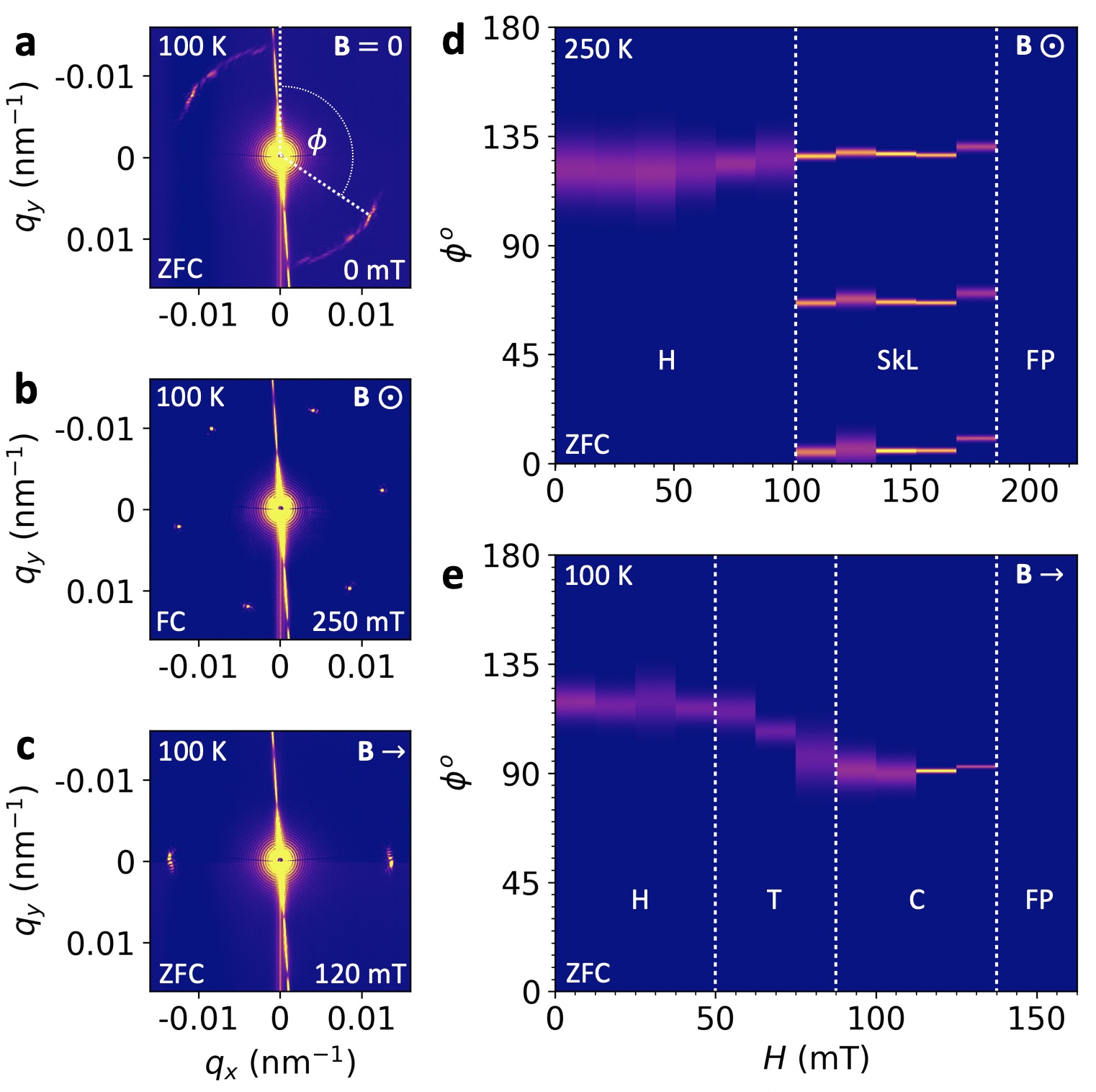}
\caption{$\vert$ \textbf{Magnetic phase diagram determination.} \textbf{a}-\textbf{c}, CCD images of the magnetic x-ray diffraction patterns obtained for the helical (H), skyrmion lattice (SkL) and conical (C) states respectively. The central yellow spot is the main x-ray beam, blocked by a beamstop, while the smaller outer peaks are the magnetic diffraction satellites. \textbf{d},\textbf{e}, Fitted angular profiles of the magnetic diffraction patterns plotted as a function of applied out-of-plane and in-plane magnetic field respectively. Determined boundaries between the magnetic states are labelled and shown by the dashed white lines.}
\label{figS3}
\end{figure}
Magnetic x-ray diffraction patterns obtained from sample 1 are shown in Fig. S3\textbf{a}-\textbf{c}. The two broad magnetic satellites in the Fig. S3\textbf{a} are characteristic of a helical state with weak alignment along a preferred crystalline axis. The six magnetic peaks in Fig. S3\textbf{b} indicate the presence of the SkL state under an out-of-plane aligned field, while the two peaks aligned to the in-plane applied field in Fig. S3\textbf{c} are representative of the conical state. By recording the CCD images as a function of applied magnetic field after ZFC, such diffraction patterns were used to plot the phase diagram in Fig. 1\textbf{a} and \textbf{b} in the main text. For each diffraction pattern, the total scattering intensity at each azimuthal angle $\phi$, as shown in Fig. S3\textbf{a}, was summed over a $q$ range from 0.01 to 0.015 nm$^{-1}$ and fitted with a Gaussian distribution. Figure S3\textbf{d} displays the fitted intensity and angular position of one of each pair of magnetic peaks as a function of applied out-of-plane magnetic field at 250 K. The broad magnetic peak of the helical state transforms to three narrow magnetic peaks in the SkL state, indicating a transition from multiple helical domains with differing orientations to a highly ordered, single SkL domain state. A similar data set for an in-plane magnetic field at 100 K is shown in Fig S3\textbf{e}, illustrating the transition of the broad helical magnetic peak through a rotation of the spin texture, to the conical state.

\section*{Location of Magnetic Skyrmion tubes}
Further STXM images are displayed in Fig. S4. The position of the observed SkT spin texture after FC from 270 K to 100 K at 35 mT is exhibited in Fig. S4\textbf{a}. The edge of the sample marked with white dotted lines indicates the limit of the magnetic region. The FeGe lamella extends for a further $\sim$200 nm beyond this line, indicating a magnetically dead layer at the boundary of the sample due to ion implantation from the fabrication process. Comparative STXM images of this corner region of the sample are shown after ZFC, revealing the local orientation of the helical state in Fig. S4\textbf{b}, and after the application of an in-plane field of 35 mT, showing formation of the conical state along the magnetic field direction. A schematic illustration of temperature-field path taken during these measurements is shown in Fig. S4\textbf{d}.

\begin{figure}
\centering
\includegraphics[width=\textwidth]{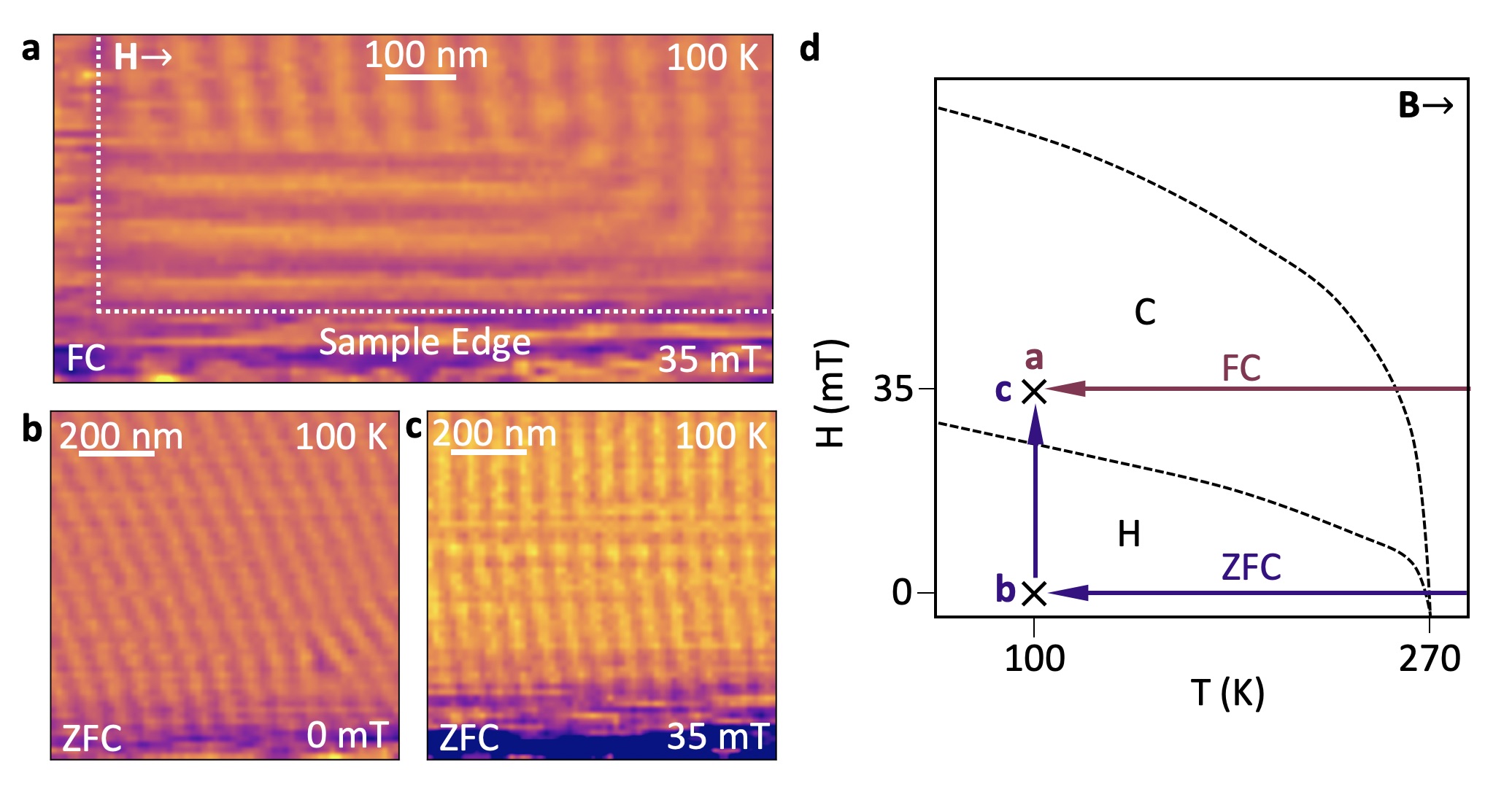}
\caption{$\vert$ \textbf{Experimental location of the magnetic skyrmion tubes.} \textbf{a}-\textbf{c}, Scanning transmission x-ray microscopy images of the same corner area of the sample recorded after FC (\textbf{a}) and ZFC (\textbf{b},\textbf{c}) procedures. The boundary of the magnetic sample is highlighted by the white dashed lines in \textbf{a}. \textbf{d}, Illustrative phase diagram depicting the-field temperature paths taken for each measurement.}
\label{figS4}
\end{figure}

\section*{Micromagnetic Simulations}
\begin{figure}
\centering
\includegraphics[width=\textwidth]{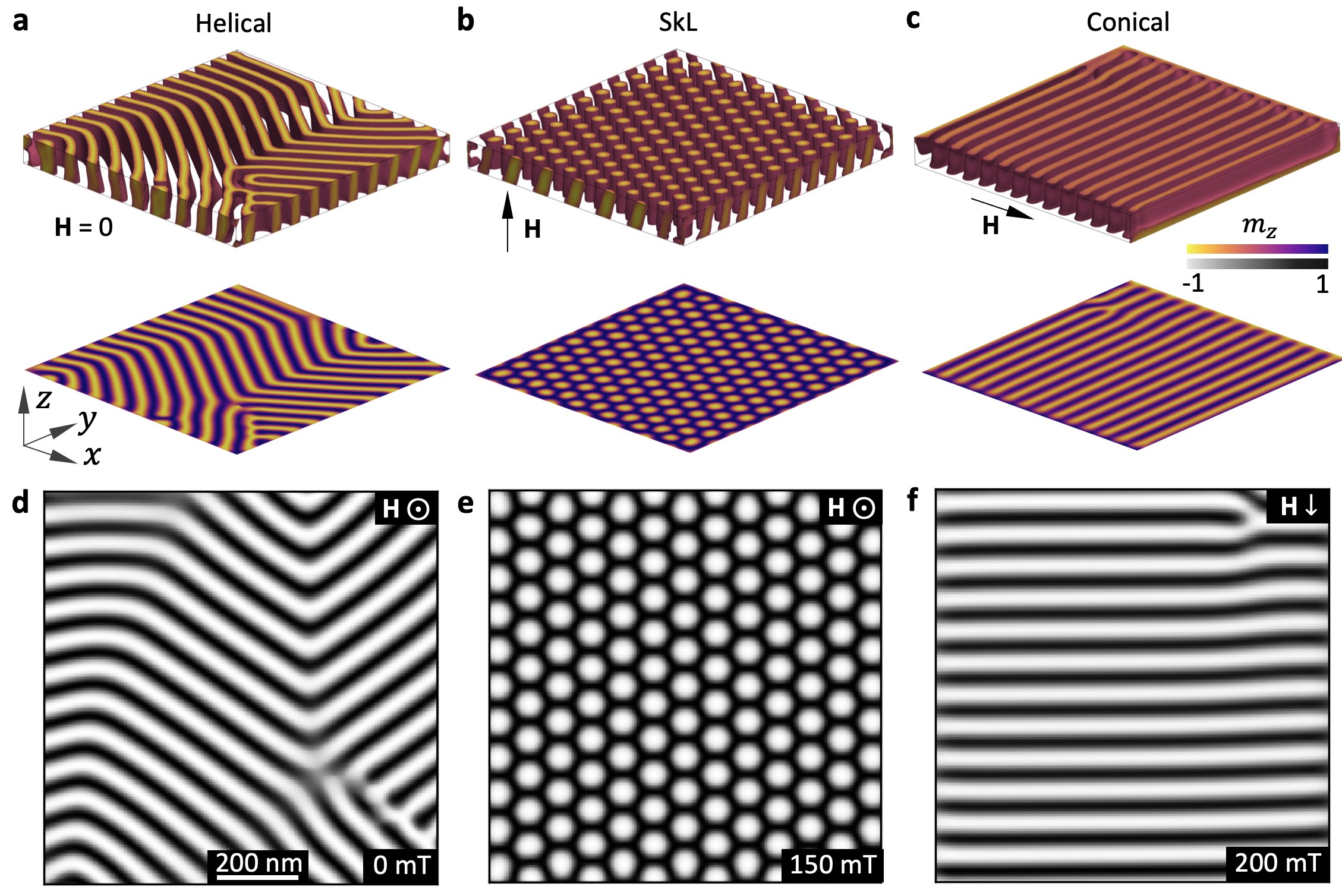}
\caption{$\vert$ \textbf{Micromagnetic simulations of chiral spin textures.} \textbf{a}-\textbf{c}, Three dimensional visualisations of the helical, SkL and conical states presented in Fig. 2 in the main text, created by plotting cells with $m_z$ components between -1 and 0. \textbf{e}-\textbf{f}, Simulated images of each simulated state, as shown in Fig. 2 of the main text.}
\label{figS5}
\end{figure}

To generate the simulated states shown in Fig. 3 of the main text, different states were initiated and then relaxed using an energy minimisation algorithm at different applied field strengths. To initialise the SkL state a triple-$q$ model was used. In the case of the conical and helical states, the simulation was initialised with one-dimensional spirals with a periodicity 70~nm. The top three dimensional visualisations in Fig.~S\ref{figS5}\textbf{a}-\textbf{c} were created by plotting cells with an out-of-plane magnetisation component $m_{z}<0$. The bottom images are $z$-slices through the centre of the simulated sample. The black and white plots in Fig.~S\ref{figS5}\textbf{d}-\textbf{f} are averages of the out-of-plane magnetisation component $m_z$ across the thickness of the sample, as shown in Fig. 3 of the main text. More details about the creation of the simulated images are discussed in the next section. 

\section*{Calculation of Simulated Images}
\begin{figure}
\centering
\includegraphics[width=0.7\textwidth]{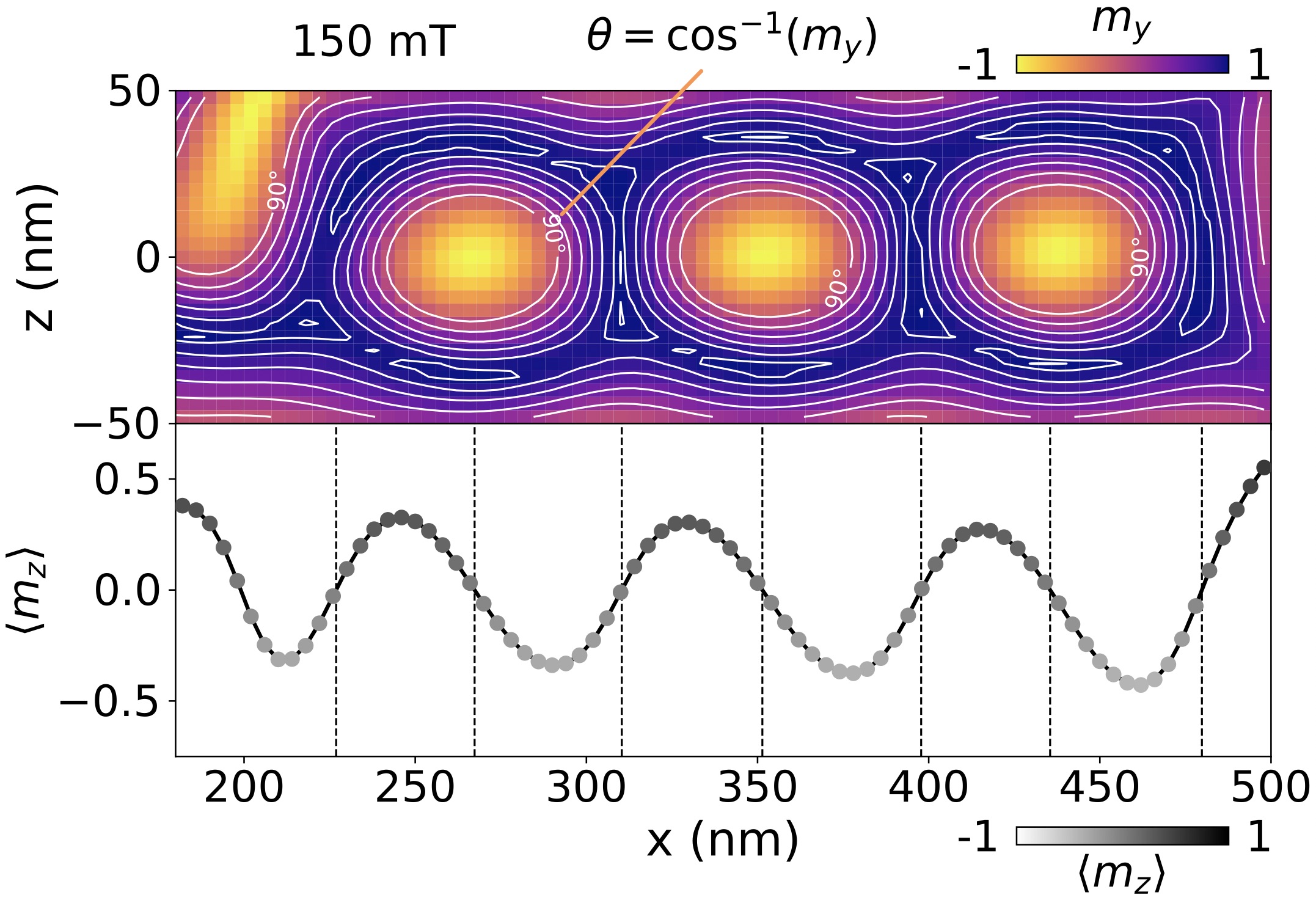}
\caption{$\vert$ \textbf{Calculating the simulated images.} The $m_y$ component of a cross section through the simulated skyrmion tube state at 150 mT is plotted in the top panel. Isocontours highlight the lines of constant $m_y$. The calculated average out-of-plane component, $\langle m_z \rangle$, is plotted in the bottom panel. Vertical dotted lines indicate positions of maximum/minimum $m_y$.}
\label{figS6}
\end{figure}
The averaging process used to create the simulated images is most clearly visualised from Fig.~S\ref{figS6}, where magnetisation profiles of a $y$-slice through the thickness of the simulated sample are shown. The top image of Fig.~S\ref{figS6} illustrates the in-plane $y$-component of the magnetisation together with isocontour lines. The $\theta=90^{\circ}$ curve refers to the $m_y=0$ surface, which encloses a skyrmion tube or a configuration pinned at the top surface, as shown at the top left of the image. The bottom plot of Fig. S\ref{figS6} shows the average $m_z$ component across the thickness of the sample ($z$-direction), $\langle m_{z} \rangle$, where the strongest contrast is the positive out-of-plane component.

Zero values of the $\langle m_{z} \rangle$ were computed using a linear interpolation of the curve and a root finding algorithm and are displayed as the dashed vertical lines. It can be seen that these zeros coincide with both the skyrmion tube centres and the middle point between two skyrmion tubes. Accordingly, the skyrmion centre zeros can be used to compute the inter-skyrmion distance and the distance of a skyrmion from the edge of the sample. In the main text the skyrmion-skyrmion and skyrmion-edge distances from Fig.~S5\textbf{b} were both computed using the zero at the centre of the rightmost skyrmion tube. In the case of skyrmion-skyrmion distance, it was computed from the zero of the adjacent skyrmion at the left.

\section*{Three Skyrmion Tube States}
\begin{figure}
\centering
\includegraphics[width=\textwidth]{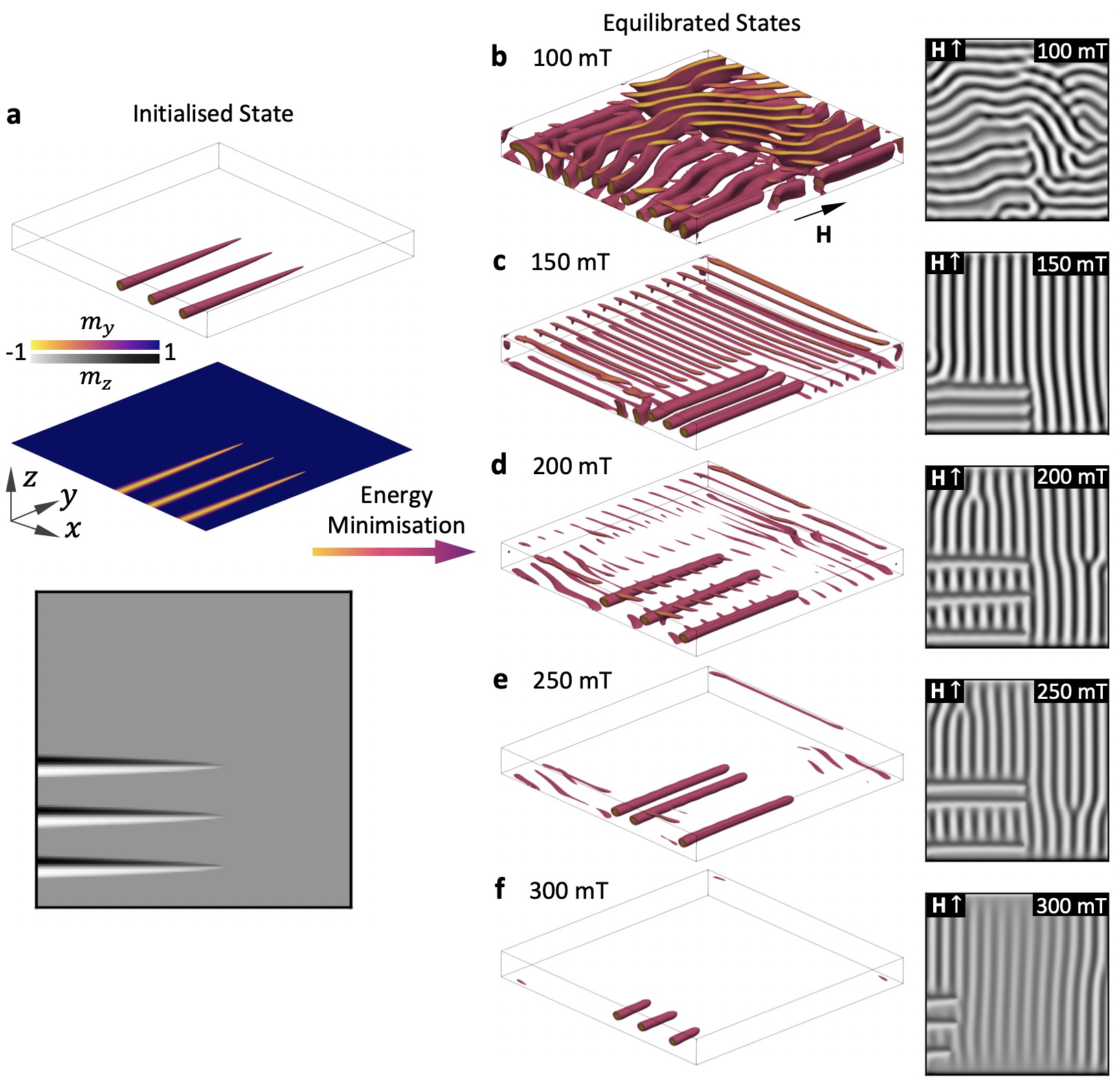}
\caption{$\vert$ \textbf{Initialisation of skyrmion tube state.} \textbf{a}, Visualisation of the initial skyrmion tube state, featuring three paraboloid-shaped skyrmion tube precursors in a ferromagnetic background. The simulated x-ray image is shown below. \textbf{b}-\textbf{f}, Visualisations, created by plotting cells with $m_y$ components between -1 and 0, and simulated x-ray images of the magnetic state after energy minimisation at a range of in-plane applied magnetic fields.}
\label{figS7}
\end{figure}
To generate the simulated three skyrmion tube states presented in the main text, a function was applied to generate three paraboloid-shaped Bloch skyrmion precursors, with the core oriented opposite to the magnetisation of the rest of the field polarised sample. By defining
\begin{equation}
\mathbf{m}=(\sin\Theta\cos\Psi,\sin\Theta\sin\Psi,-\cos\Theta),  
\end{equation}
the functions $\Theta=\Theta(r,z)$, $\Psi=\Psi(\varphi)$, with $(r,\varphi,z)$ as cylindrical coordinates, are specified as
\begin{equation}
    \begin{split}
        \Theta(r,z) & = \frac{\pi}{r_{\text{par}}(z)} r,\quad 
        r_{\text{par}}(z) = \sqrt{\frac{r_{\text{tube}}}{z_{\text{tube}}} z},\quad 0 \leq z \leq 600\,\text{nm} \\
        \Psi(\varphi) & = \varphi + \frac{\pi}{2}.
    \end{split}
\end{equation}
\noindent Based on these functions, each of the three tubes was initialised with $r_{\text{tube}}=40\,\text{nm}$ and $z_{\text{tube}}=600\,\text{nm}$. The largest radius of the tube was specified with a value smaller than the helical length of FeGe of because of the limited thickness of the 100 nm sample. The tube length was chosen according to the length of the tubes observed from the experimental images. To generate three tubes, the first tube was located with its centre at a distance of 125~nm from the edge of the sample and the second and third tubes were positioned with a 160~nm separation. The rest of the sample was initialised in a field-polarised state with magnetisation $\mathbf{m}=(0, +1, 0)$. 

An overview of the initial state for the three tube system is shown in Fig.~S\ref{figS7}\textbf{a}. The regions within the $m_y=0$ isosurfaces are shown in the top image and a $z$-slice at the centre of the sample is shown at the middle image, together with a simulated image at the bottom. This initial state was relaxed at a range of magnetic fields by minimising the energy using a conjugate gradient algorithm implemented in the OOMMF code\cite{OOMMFs}, and the resulting states are are shown in Fig.~S\ref{figS7}\textbf{b}-\textbf{f}. The 150 mT simulated state most resembles the three tube configuration seen in the experiments, displayed in Fig.~S\ref{figS7}\textbf{c}. Therefore, we chose this state to simulate the field sweep and observe the evolution of the skyrmion tubes with increasing and decreasing magnetic fields for comparison with the experimental data. 

\section*{Randomly Initialised Field Sweeps}
\begin{figure}
\centering
\includegraphics[width=0.55\textwidth]{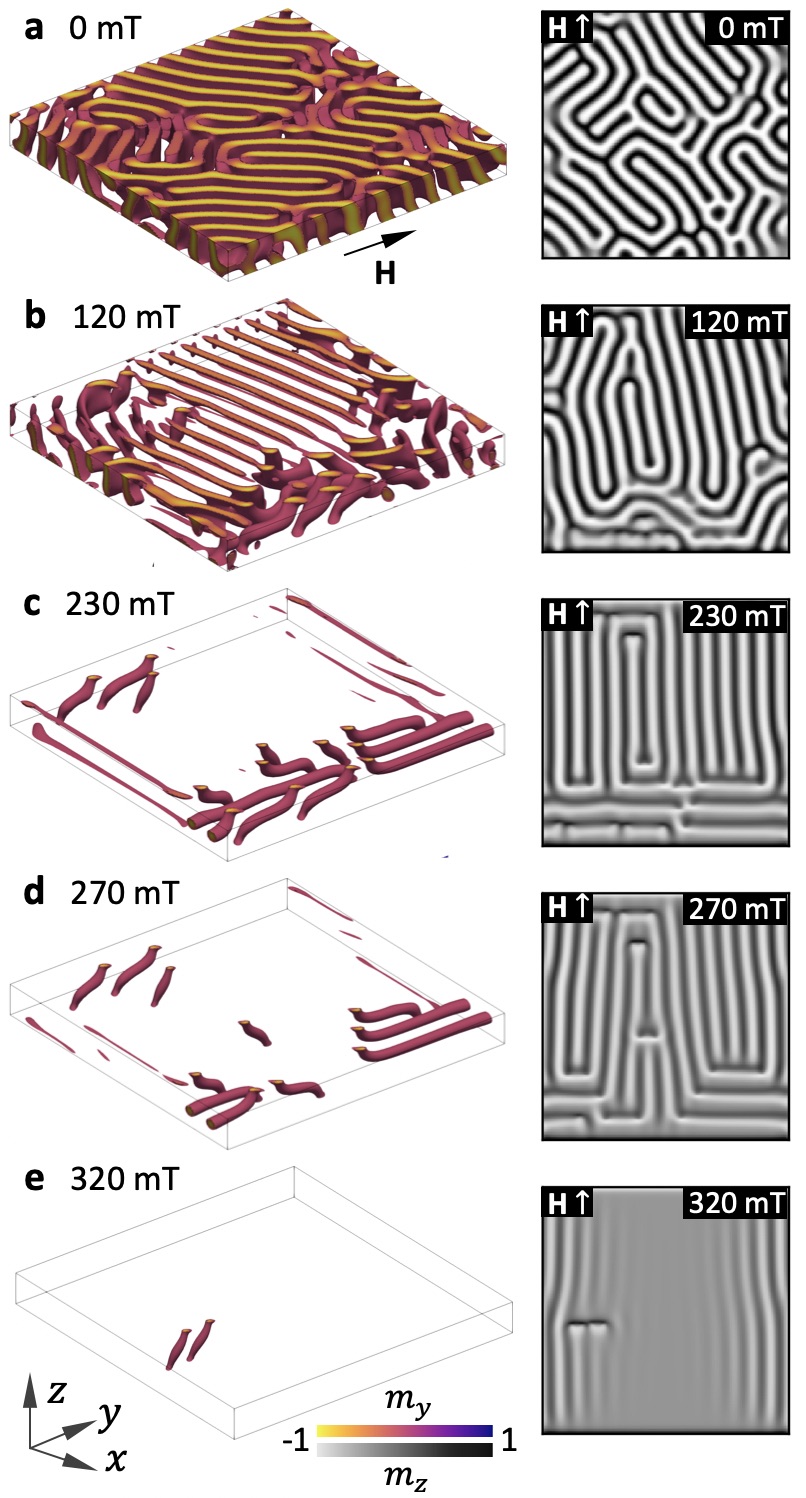}
\caption{$\vert$ \textbf{In-plane magnetic field simulation from random initial state.} \textbf{a}-\textbf{e}, Selected three dimensional visualisations, created by plotting cells with $m_y$ components between -1 and 0, and simulated images of the magnetic state during an in-plane field sweep from an initially randomised state.}
\label{figS8}
\end{figure}
\begin{figure}
\centering
\includegraphics[width=0.55\textwidth]{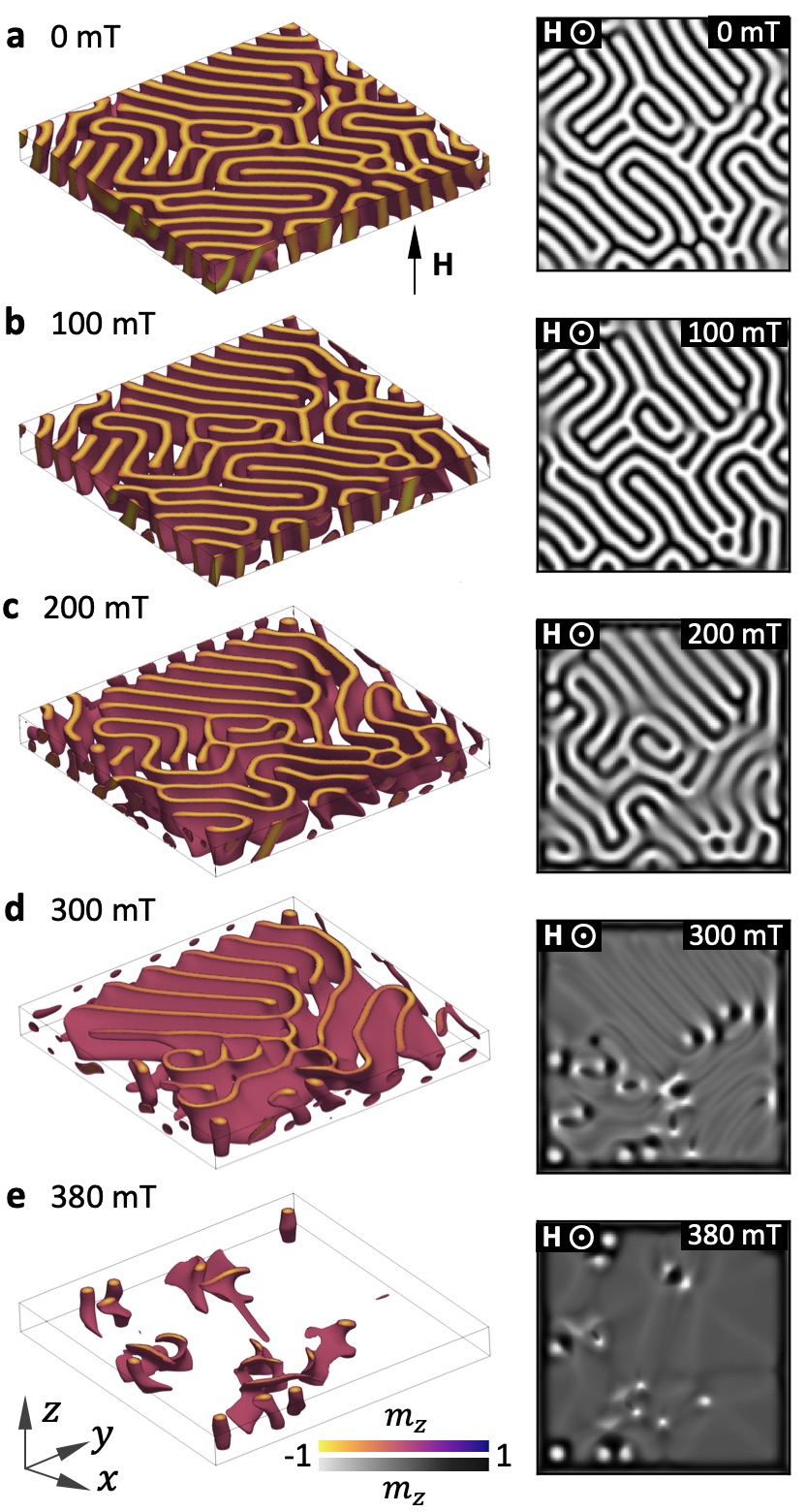}
\caption{$\vert$ \textbf{Out-of-plane magnetic field simulation from random initial state.} \textbf{a}-\textbf{e}, Selected three dimensional visualisations, created by plotting cells with $m_z$ components between -1 and 0, and simulated images of the magnetic state during an out-of-plane field sweep from an initially randomised state.}
\label{figS9}
\end{figure}
In order to examine the possibility of skyrmion tube formation from a field sweep starting at 0 mT, we initialised a helical state at 0 mT from a fully randomised precursor state, performing an energy relaxation using the conjugate gradient minimisation method. The resulting initial state is shown in Fig.~S\ref{figS8}\textbf{a}, with $m_y$ isosurfaces in the left plot and the average out of plane magnetisation through the thickness, $\langle m_{z} \rangle$, in the right plot. Two field sweep processes were simulated starting from this initial zero field helical state, applying both an in-plane and an out-of-plane magnetic field. Results of the in-plane field sweep are shown in Fig.~S\ref{figS8}\textbf{a}-\textbf{e}. Images at the left side of these figures show the region within the $m_y=0$ isosurface. As the magnetic field is increased skyrmion tubes form with their ends pinned at the sample surfaces, which may be energetically favoured in comparison to the formation of Bloch points. Results of the out-of-plane field sweep are shown in Fig.~S\ref{figS9}\textbf{a}-\textbf{e}, with $m_z$ isosurfaces depicted in the images to the left side of every field sweep stage. In this case, skyrmions are again primarily stabilised close to the boundaries of the sample, highlighting the importance of confinement and edge effects.

\section*{Details of the Skyrmion Tube and Bloch Point State}
\begin{figure}
\centering
\includegraphics[width=0.50\textwidth]{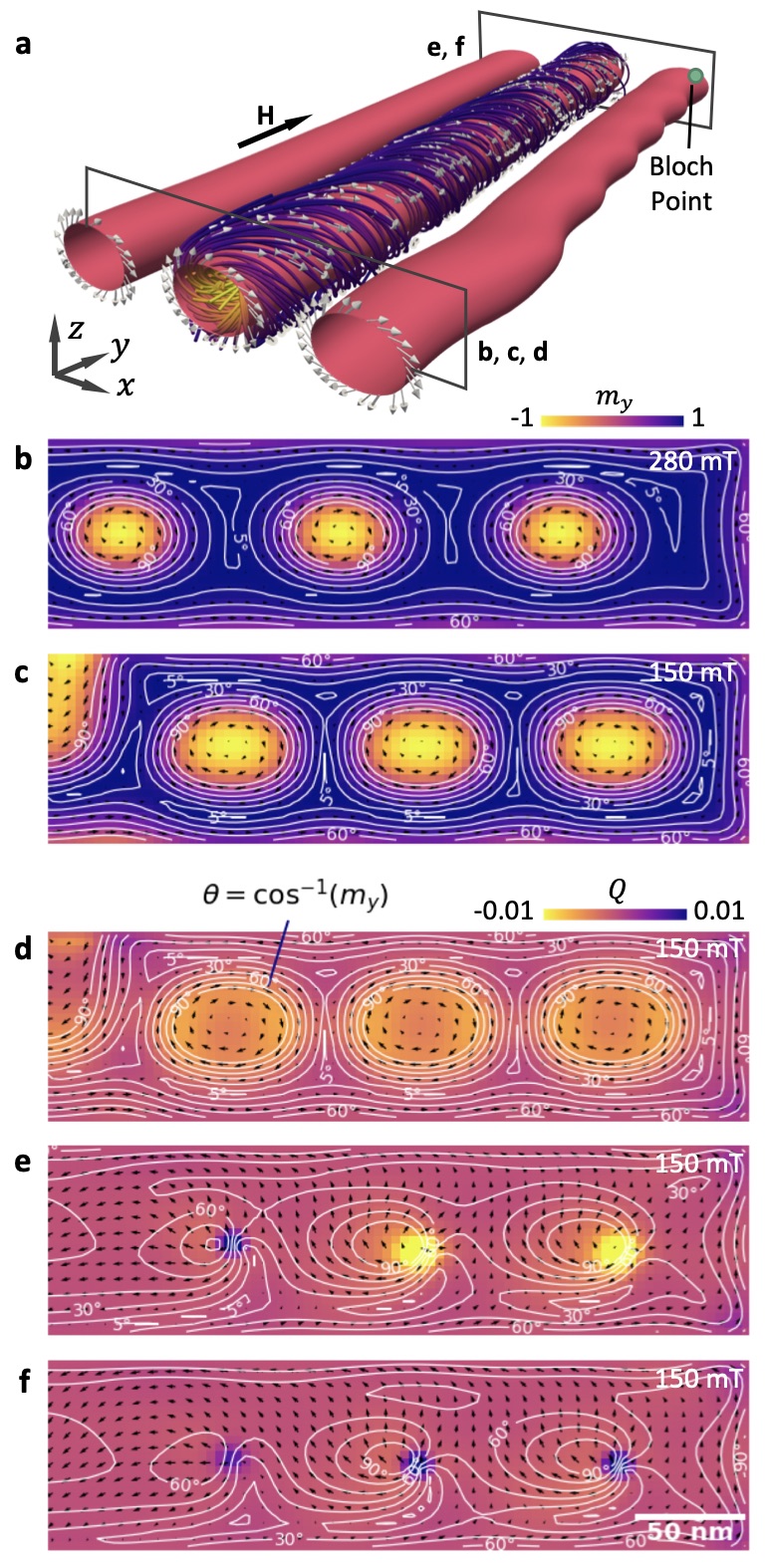}
\caption{$\vert$ \textbf{Isocontours and topological charge of skyrmion tubes.} \textbf{a}, Three dimensional visualisation of the skyrmion tube spin texture at 150 mT. \textbf{b},\textbf{c}, The $m_y$ component of cross sections through the simulated skyrmion tube states at 280 and 150 mT. Isocontours exhibit the lines of constant $m_y$. \textbf{d}-\textbf{f}, Calculated topological charge density of cross sections through the simulated skyrmion tube state at 150 mT close to the edge of the sample (\textbf{d}), and at the position of the magnetic Bloch point (\textbf{e},\textbf{f}).}
\label{figS10}
\end{figure}

A three dimensional representation of the simulated three skyrmion tubes state at 150~mT is shown in Fig.~S\ref{figS10}\textbf{a}, highlighting two cross sections through the width of the tubes, one near the edge of the simulated sample, and one at the location of the magnetic Bloch points. In Fig.~S\ref{figS10}\textbf{b} and~\textbf{c}, cross sections close to the surface of the tubes at 280 and 150~mT are shown. White lines specify isocontours of the simulated magnetisation with a constant angle $\theta$, which is measured with respect to the $m_{y}=1$ value, \textit{i.e.} $\cos\theta = m_{y}$. Isocontour surfaces have been used previously to show the asymmetric character of isolated skyrmion tubes embedded within a conical background, and highlight the Lennard-Jones like attraction when two adjacent tubes form a coupled state with overlapping isocontours\cite{leonov_three_2016s,du_interaction_2018s}, reaching an equilibrium distance. At higher applied magnetic fields, the background becomes field polarised, and the skyrmion tube isocontours no longer distort, causing two adjacent skyrmions to repel\cite{du_interaction_2018s}. In the case of the confined skyrmion tubes analysed here, the observed structures are more complex as they end in a Bloch point singularity. Within the conical phase, besides the skyrmion tubes being slightly distorted along their lengths, there is no evidence of asymmetric isocontours surfaces, possibly due to the constrained sample dimensions. Despite this, the repulsion between neighbouring skyrmions and the sample edge with increasing magnetic field is still observed, in agreement with past calculations. 
A profile of the topological charge density $q$ of the surface cross section at 150~mT, is shown in Fig.~S\ref{figS10}\textbf{d}. The density of topological charge, or skyrmion number, is defined as 
\begin{equation}
        q = \frac{1}{4\pi} \mathbf{m} \cdot \left( \frac{\partial
        \mathbf{m}}{\partial z} \times \frac{\partial \mathbf{m}}{\partial x}\right).
    \label{eq:skyrmion-number} 
\end{equation}
In the cross section close to the surface of the skyrmion tube, shown by Fig.~S\ref{figS10}\textbf{d}, the topological charge density is given by the non-colinearity of the spins around the $\theta=90^\circ$ region, and this is repeated through the skyrmion length. Towards the end of the skyrmion tubes, at the location of the magnetic Bloch points, the $\theta=90^\circ$ area is reduced and $q$ is concentrated in a infinitesimal region. Interestingly, just before and after the end of each skyrmion tube, the polarity of the charge density reverses from negative to positive, as shown by Figs.~S\ref{figS10}\textbf{e} and \textbf{f} respectively, revealing the singular character of the magnetic Bloch point. The simulated system is discretised with cuboid cells of 4~nm side length, hence it is likely that these Bloch point regions are not well defined and the energy cost of forming such a state may be underestimated. A better definition of this magnetisation requires an atomistic description, which is nontrivial due to the large size of the simulated sample. 

\section*{Distortion of the skyrmion tube magnetisation profile}
To characterise the distortion of the magnetisation profiles of the skyrmion tubes, we use the mathematics of the distortion of the helical state into a chiral soliton lattice with applied field\cite{Dzyaloshinskii1965,Izyumov1984}. We can write,
\begin{equation}
\frac{\text{E}(\kappa)}{\kappa} = \sqrt{\frac{1}{h}},
\end{equation}
where $\text{E}(\kappa)$ is the complete elliptical integral of the second kind, $h$ is the ratio between the applied field and the critical field ($H / H_c$), and $\kappa$ is the modulus of the elliptical function, for which we want to solve. Using the value of $\kappa$ determined by minimising this equation, the magnetisation profile of the distorted helial state for the vector $\vec{Q} \parallel \hat{x}$ and $\vec{H} \parallel \hat{z}$ is,
\begin{equation}
\begin{split}
\vec{M} & = M_s(0,M_y,M_z), \\
M_y & = \sin(2\text{am}(\sqrt{H}\frac{x}{\kappa},\kappa)), \\
M_z & = \cos(2\text{am}(\sqrt{H}\frac{x}{\kappa},\kappa)),
\end{split}
\end{equation}
where $\text{am}(u,v)$ is the Jacobi amplitude function, and $M_s$ is the saturation magnetisation. This can be conveniently expressed using the Jacobi Elliptic functions $\text{sn}(u,v) = \sin(\text{am}(u,v))$ and $\text{cn}(u,v) = \cos(\text{am}(u,v))$ as,

\begin{equation}
\begin{split}
M_y & = 2 \text{cn}(\sqrt{H}\frac{x}{\kappa},\kappa)\text{sn}(\sqrt{H}\frac{x}{\kappa},\kappa), \\
M_z & = \text{cn}^2(\sqrt{H}\frac{x}{\kappa},\kappa) - \text{sn}^2(\sqrt{H}\frac{x}{\kappa},\kappa).
\label{eq:magprof}
\end{split}
\end{equation}

\begin{figure}
\centering
\includegraphics[width=0.9\textwidth]{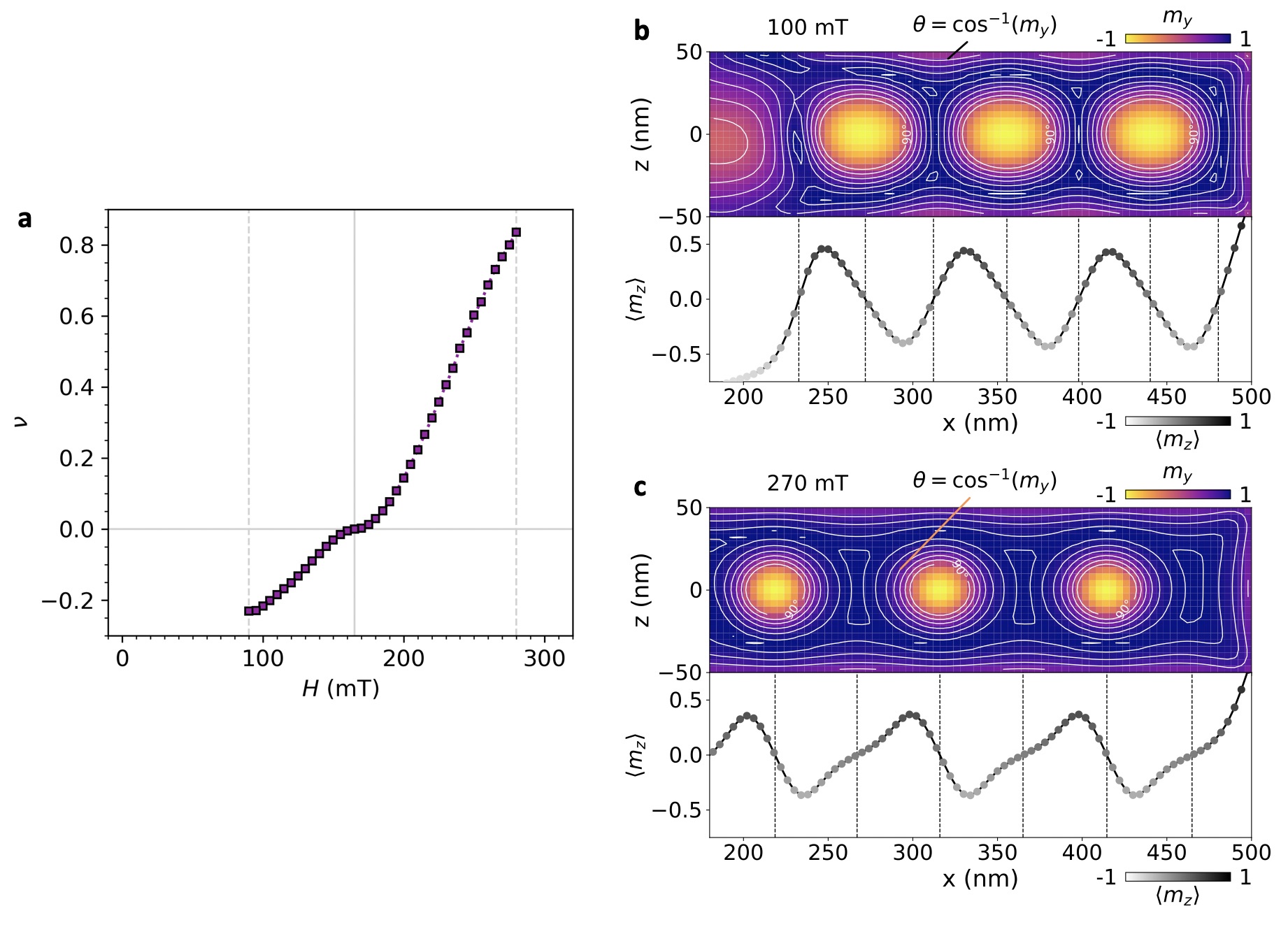}
\caption{$\vert$ \textbf{Distortion of the skyrmion tube magnetisation profiles.} \textbf{a}, The parameterised distortion, $\nu$, of the out-of-plane magnetisation profile of the skyrmion tube state as a function of applied magnetic field. \textbf{b},\textbf{c}, The $m_y$ component of cross sections through the simulated skyrmion tube states at 100 and 270 mT are plotted in the top panels. Isocontours highlight the lines of constant $m_y$. The calculated average out-of-plane component, $\langle m_z \rangle$, is plotted in the bottom panels. Vertical dotted lines indicate positions of maximum/minimum $m_y$.}
\label{figS11}
\end{figure}

For the magnetisation profiles the relevant component of the magnetisation is $M_y$. This expression is then used to fit the magnetisation profiles, using the effective applied field $h$ as a fitting parameter, and adding in an adjustable phase offset $\phi$ such that $x \rightarrow x+\phi$. This yields,

\begin{equation}
M_y  = 2 \text{cn}(\sqrt{H}\frac{x+\phi}{\kappa},\kappa)\text{sn}(\sqrt{H}\frac{x+\phi}{\kappa},\kappa). \\
\end{equation}

For the high field curves above 165 mT, this equation fits the simulated magnetization profiles. For the lower field curves, the distortion is seen to have the opposite directionality. This can be mathematically described by adding a negative sign and a phase offset of $\pi$, giving,
 
 \begin{equation}
M_y  = -2 \text{cn}(\sqrt{H}\frac{x+\phi-\pi}{\kappa},\kappa)\text{sn}(\sqrt{H}\frac{x+\phi-\pi}{\kappa},\kappa). \\
\end{equation}

As shown in Fig.~S\ref{figS11}, the out-of-plane component of the magnetisation across the width of the SkT exhibits an asymmetrical distortion that varies as a function of applied magnetic field and is not observed in the conical state. The ratio of the applied field to the critical field, $h$, parameterises the magnitude of this distortion. We define $\nu=h$ for fields of 165 mT and above, and $\nu=-h$ for fields less than 165 mT, where the SkT state is essentially undistorted. This can be seen in Fig.~S\ref{figS11}\textbf{a}, where $\nu$, obtained by fitting the magnetisation profiles with the chiral soliton lattice equations, is plotted as a function of applied magnetic field. The profiles of the SkT state reveal that the direction of the asymmetry is opposite at low and high magnetic fields. This is most easily visualised in Fig.~S\ref{figS11}\textbf{b} and \textbf{c}. At low applied fields, such as at 100 mT in Fig.~S\ref{figS11}\textbf{b}, the magnetisation profile varies fastest between the skyrmion tubes, due to the reduced separation of the tubes relative to their radii. Conversely, at high fields, such as at 270 mT in Fig.~S\ref{figS11}\textbf{c}, the magnetisation profile changes fastest within each skyrmion tube, due to the increased separation of the tubes relative to their radii. This leads to the magnetisation profile of the tubes appearing distorted, due to the varying skyrmion tube separation.

\end{document}